\documentclass{aastex631}
\usepackage{amsmath}

\shorttitle{Acceleration and Feedback of Electrons in Solar Flares}
\shortauthors{Seo et al.}

\begin{document}

\title{Energetic Nonthermal Electrons within the Above-the-looptop Regions in Solar Flares: Acceleration, Feedback, and Quasi-Periodic Pulsations}
\author[0000-0002-5550-8667]{Jeongbhin Seo}
\affiliation{Los Alamos National Laboratory, Los Alamos, NM 87545, USA}
\author[0000-0003-4315-3755]{Fan Guo}
\affiliation{Los Alamos National Laboratory, Los Alamos, NM 87545, USA}
\affiliation{New Mexico Consortium, Los Alamos, NM 87544, USA}
\author[0000-0001-5278-8029]{Xiaocan Li}
\affiliation{Los Alamos National Laboratory, Los Alamos, NM 87545, USA}
\author[0000-0002-0660-3350]{Bin Chen}
\affiliation{Center for Solar-Terrestrial Research, New Jersey Institute of Technology, 323 Dr. Martin Luther King Jr. Boulevard, Newark, NJ 07102-1982, USA}
\author[0000-0002-9258-4490]{Chengcai Shen}
\affiliation{Harvard-Smithsonian Center for Astrophysics, 60 Garden Street, Cambridge, MA, 02138, USA}
\author[0000-0003-3556-6568]{Hui Li}
\affiliation{Los Alamos National Laboratory, Los Alamos, NM 87545, USA}

\begin{abstract}
Solar flares are among the most dramatic events in the solar system, releasing substantial magnetic energy and accelerating a large number of electrons to high energies. Notably, in certain events, the above-the-looptop region may contain a significant population of nonthermal electrons, both in number and energy. For the first time, we adopt a novel numerical method that combines magnetohydrodynamics with energetic particles incorporating feedback from nonthermal electrons to investigate electron acceleration and transport in solar flares. We find that
a large fraction of energetic electrons are accelerated via the current sheet and termination shock regions. Most energetic electrons are concentrated in the above-the-looptop region, carrying a sizable amount of the released energy. 
 We observe that greater feedback of nonthermal electrons leads to steeper energy spectra.  The energy density of the nonthermal electrons oscillates due to the periodic impact of magnetic islands into the above-the-looptop, which may help explain the observed quasi-periodic pulsations (QPPs). Our simulations provide new insights into the origin of nonthermal electrons and associated emissions in the above-the-looptop region.

\end{abstract}

\section{Introduction}
During solar flares, magnetic reconnection drives rapid release of a substantial amount of magnetic energy \citep{carmichael1964,sturrock1966,hirayama1974,kopp1976}, resulting in strong plasma heating and nonthermal particle acceleration. A significant portion of the released energy is believed to be carried by nonthermal electrons \citep[e.g.,][]{kruker2010,oka2013looptop,oka2015,warmuth2020}, and understanding the mechanisms behind their acceleration remains a central focus of solar flare research.

The multi-wavelength emissions (hard X-rays and microwaves), generated by these accelerated electrons, provide valuable insights into particle acceleration, transport, and radiation processes. These observations have revealed that nonthermal electrons typically exhibit a power-law energy spectrum, $f(\varepsilon) \propto \varepsilon^{-\delta}$, with a median spectral index of $\delta \sim 5$ \citep{kruker2010,oka2013looptop,oka2015,Effenberger2017}.  Hard X-ray observations suggest that, in certain events, a significant fraction of electrons are accelerated into the nonthermal range, carrying energy comparable to the released magnetic energy \citep{kruker2010,oka2013looptop,oka2015}. Especially, recent EOVSA observations along with other multiwavelength observations have provided unprecedented imaging spectroscopy of the X8.2-class eruptive flare event SOL2017-09-10 \citep{gary2018,Chen2020,fleishman2022}. These observations show that the nonthermal electrons are partially confined to the above-the-looptop region with a relatively weak magnetic strength, referred to as the ``magnetic bottle'' \citep{kong2019,Chen2020, Chen2024}. It has been suggested that the above-the-looptop region is the primary region for energetic particle acceleration and trapping in this event, although further studies are needed to see if this is a universal scenario. 

\citet{fleishman2022} also presented spatially resolved distributions of thermal and nonthermal electrons in the SOL2017-09-10 flare, based on the analysis of EOVSA data. These distributions suggested that the above-the-looptop region is predominantly filled with nonthermal electrons, almost devoid of thermal plasma. It is worth noting that \citet{kontar2023} studied the same flare and concluded that the fraction of nonthermal electrons in the above-the-looptop region was much lower than that estimated by \citet{fleishman2022}.
Nevertheless, numerous studies have suggested a large fraction of the accelerated electrons in the flare region are nonthermal \citep{krucker2010,oka2013looptop,oka2015,fleishman2022}. These suggest that feedback from nonthermal electrons may play an important role in the above-the-looptop region.

Studying such efficient electron acceleration in solar flares requires new methods to capture the acceleration processes, the large-scale flare dynamics, and the potential feedback from energetic electrons to the background plasmas. Recently, we developed a novel method for simulating particle acceleration with feedback from energetic particles, called the MHD-SDE method \citep{seo2024feedback}. This method incorporates feedback from the pressure of nonthermal particles in MHD simulations, obtained by solving Parker's transport equation \citep{parker1965} using the stochastic differential equation (SDE) method. The Parker transport equation includes the spatial diffusion of nonthermal particles by MHD turbulence, allowing for more realistic simulation of transport and acceleration. 

For the first time, we apply the MHD-SDE method to a flare eruption geometry, focusing on the above-the-looptop region where nonthermal electron numbers and energies are prominent. By investigating nonthermal electron distributions and acceleration mechanisms in this region, we aim to enhance our understanding of solar flare dynamics and the impact of energetic particle feedback.
Our results suggest that nonthermal electrons, accelerated along the current sheet and at the termination shock, follow a power-law energy spectrum whose slope varies depending on the feedback of the nonthermal electron. The impact of magnetic islands into the above-the-looptop generates oscillations in the energy density of nonthermal electrons, which may lead to the generation of quasi-periodic pulsations (QPPs).
The paper is organized as follows: Section \ref{s2} introduces the numerical method, followed by a discussion of the simulation results in Section \ref{s3}. A brief summary and outlook are provided in Section \ref{s4}. 

\section{Numerical Method}
\label{s2}
\subsection{MHD-SDE model}

Several approaches have been developed to incorporate the feedback of nonthermal component at the hydrodynamic scale \citep[see][for a review]{ruszkowski2023}. A common approach is to include feedback effects of energetic particles in the MHD equations. Examples include the $\it{kglobal}$ model \citep{drake2019,Arnold2021}, the MHD-PIC method \citep{bai2015,sun2023}, the non-linear kinetic Monte Carlo model \citep{bykov2014,bykov2024}, and the turbulence transport equation coupled with the MHD equation \citep{wang2022}. However, none of these methods have been applied to solar flares with a realistic magnetic geometry.

For the first time, we have taken into account the feedback from nonthermal electrons in an eruptive flare geometry. These electrons are accelerated and carry a significant fraction of the released energy. 
We have developed the MHD-SDE method to simulate systems where the pressure from nonthermal particles is significant \citep{seo2024feedback}. This method solves the MHD equations with feedback from the nonthermal particles. To implement this feedback, we obtain the nonthermal pressure from the distribution function of the nonthermal component, which is derived from Parker's transport equation using stochastic differential equation (SDE). 

The resistive MHD equations with nonthermal fluid feedback are derived as \citep{seo2024feedback}

\begin{eqnarray}
    \frac{\partial \rho}{\partial t} + \nabla\cdot(\rho \boldsymbol{v}) = -S_\rho\label{cont},\\
    \frac{\partial\rho\boldsymbol{v}}{\partial t} + \nabla\cdot\left(\rho \boldsymbol{v}\boldsymbol{v} + P^*\boldsymbol{I}-\boldsymbol{B}\boldsymbol{B}\right) = -\nabla \cdot P_{\rm{NT}}\boldsymbol{I}-S_m, \label{meq}\\
    \frac{\partial E}{\partial t} + \nabla\cdot\left[(E+P^*)\boldsymbol{v}-\boldsymbol{B}(\boldsymbol{B}\cdot \boldsymbol{v})\right] = \nabla\cdot(\boldsymbol{B}\times\eta\boldsymbol{j})-\boldsymbol{v}\cdot(\nabla \cdot P_{\rm{NT}}\boldsymbol{I})-S_E, \label{Eeq}\\
    \frac{\partial \boldsymbol{B}}{\partial t} - \nabla\times(\boldsymbol{v}\times\boldsymbol{B}) = \eta\nabla^2\boldsymbol{B}, \label{indu}
\end{eqnarray}
where
\begin{eqnarray}
    E=P_g/(\gamma_g-1) + (\rho \boldsymbol{v}\cdot\boldsymbol{v} + \boldsymbol{B}\cdot\boldsymbol{B})/2.
\end{eqnarray}
Here, $\rho$, $\boldsymbol{v}$, $\boldsymbol{B}$, $P_g$, $P_{\rm NT}$, $\boldsymbol{j}$, $\gamma_g$, and $\eta$ denote the thermal plasma density, velocity, magnetic field, thermal plasma pressure, nonthermal electorn pressure, current density, adiabatic index of the thermal plasma, and resistivity, respectively. $P^*$ is the sum of thermal plasma pressure and magnetic pressure, $P^* = P_g + B^2/2$. 

$S_\rho$, $S_m$, and $S_E$ are the sink terms due to nonthermal particle injection from thermal plasmas for density, momentum, and energy, respectively, and will be discussed in Section \ref{s2.3}. In this context, the subscripts $g$ and ${\rm{NT}}$ denote the thermal plasma gas and the nonthermal electrons, respectively.
Note that our current model omits the effects of thermal conduction, which we defer to a future study. This omission is not unique to our work \citep[e.g.,][]{Dahlin2025}, as incorporating all the multiphysics effects (including energetic particles, the focus of the current study) during solar flares remains very demanding.

We solve Parker's transport equation for nonthermal electrons to obtain their distribution as follows
\begin{equation}
\frac{\partial f}{\partial t}+(\boldsymbol{v}+\boldsymbol{v_d})\cdot\nabla f-\frac{1}{3}\nabla \cdot \boldsymbol{v} \frac{\partial f}{\partial \ln p_{\rm{NT}}} = \nabla\cdot(\boldsymbol{\kappa}\nabla f) + Q\label{parker}
\end{equation}
where $f(x_{\rm{NT}},p_{\rm{NT}},t)$ is the distribution function, which is a function of position, $x_{\rm{NT}}$, momentum, $p_{\rm{NT}}$, and time, $t$. $\boldsymbol{v_d}$ and $\boldsymbol{\kappa}$ are the particle drift velocity and the spatial diffusion tensor, respectively. In our simulation, we only consider cases with $\boldsymbol{v_d}$ is negligible, which is a good approximation for solar flare reconnection (e.g., $\boldsymbol{v}_d \sim$ 0.02 km s$^{-1}$ for a 10 keV electrons, current sheet thickness of 100 km, and a magnetic field of 100 G \citep{isenberg1979}). 
$Q$ is the source term with the injection distribution $s(p_{\rm{NT}})$, which will be discussed in Section \ref{s2.3}.
We solve this equation using SDE; the detailed method is described in \citet{li2018}.
To incorporate the feedback from nonthermal particles into the MHD equations, we treat the bulk population of nonthermal electrons as a fluid. 

The pressure from the nonthermal fluid is given by
\begin{eqnarray}
   P_{\rm NT} 
   = \frac{1}{3}
     \int v_{\rm NT}\, p_{\rm NT}\,
     f(x_{\rm NT}, p_{\rm NT}, t)\,
     d^3 p_{\rm NT},
   \label{relP}
\end{eqnarray}
where the speed of nonthermal particles is
\[
v_{\rm NT}
= \frac{p_{\rm NT} c^2}
        {\sqrt{p_{\rm NT}^2 c^2 + m_e^2 c^4}},
\]
where $m_e$ and $c$ are the electron mass and the speed of light, respectively. This form is applicable for particles with energies from nonrelativistic to relativistic regimes.

%The energy density of this nonthermal fluid, $E_{\rm{NT}}$,  at any given time and location is calculated by collecting the second-order moment of nonthermal electrons within each cell,
%\begin{eqnarray}
%   E_{\rm{NT}} = \int E(p_{\rm NT}) f(x_{\rm NT},p_{\rm NT},t) dp_{\rm NT}^3  \label{relP},%\rho_{\rm{NT}}\left<p_{\rm{NT}}\right>c 
%\end{eqnarray}
%where $E(p_{\rm NT})=p^2/(2m_e)$ is the particle energy as a function of its momentum. In our simulations, we assume a nearly isotropic momentum distribution for the nonthermal fluid, $p_{\rm NT}$. The nonthermal fluid is approximated to be an ideal gas with a constant specific heat ratio, and the pressure of the nonthermal electrons can be expressed in terms of their energy density \citep{kudoh2016,seo2024feedback},
%\begin{eqnarray}
%    P_{\rm{NT}} = (\gamma_{\rm{NT}}-1)E_{\rm{NT}}.
%\end{eqnarray}
%where $\gamma_{\rm{NT}}$ is the adiabatic index for the nonthermal fluid. This index is 5/3 for nonrelativistic electrons. \textcolor{blue}{For simplicity, we adopt a single nonrelativistic adiabatic index for the electrons; therefore, we consider only electrons with energies below 500 keV in our simulations.}

The Parker transport equation assumes a nearly isotropic momentum distribution due to frequent pitch-angle scattering, thereby incorporating the effects of spatial diffusion and compression acceleration of energetic particles. This assumption is well supported by numerical studies. For example, first-principle Particle-in-cell (PIC) simulations have shown that compression energization is critical for accelerating high-energy particles in reconnection with a weak guide field, where particle distributions remain nearly isotropic and shear acceleration is inefficient due to low anisotropy \citep{li2018a}. MHD simulations further suggest that reconnection layers are compressible, particularly at low plasma $\beta$ and with weak guide fields \citep{birn2012,provornikova2016}; in eruptive flare simulations, strong guide fields can weaken significantly during peak reconnection \citep{dahlin2022}, and weak guide field models better match observed magnetic structures in the SOL2017-09-10 flare \citep{Chen2020}.

Earlier studies have also shown  that turbulence generated during 3D reconnection can efficiently scatter high-energy particles, leading to nearly isotropic distributions \citep{li2019a}.
In addition, energetic electrons can excite plasma waves and associated structures that act to scatter and confine them \citep[e.g.,][]{li2013,Ma2023}. Observational evidence supports this picture. For instance, \citet{chen2020ApJ} reported footpoint microwave emissions delayed by nearly 20s compared to flux-rope sources, much longer than the sub-second travel time expected for 100 keV electrons under free streaming. Similarly, \citet{Mondal2024} showed that assuming free streaming leads to an order-of-magnitude overestimate of footpoint fluxes. Together, these findings underscore the essential role of scattering in nonthermal electron transport.

In our study, the confinement of energetic electrons is due to spatial diffusion as particles are scattered in the looptop region (see more discussion below).      
It is worth-noting that the Parker transport equation does not explicitly include effects with pitch-angle dependence, such as the magnetic mirroring. Some studies have investigated electron acceleration in solar flare regions by solving the focused transport equation which includes pitch-angle dependence and magnetic mirroring, but they yield a similar nonthermal electron distribution compared to that obtained from the Parker equation \citep{kong2019,kong2022}. Further studies will be required to better understand the mechanism for confining energetic electrons at the looptop regions.

\subsection{Simulation setup}
\begin{figure}[ht!]
\centering
\includegraphics[width=0.8\textwidth]{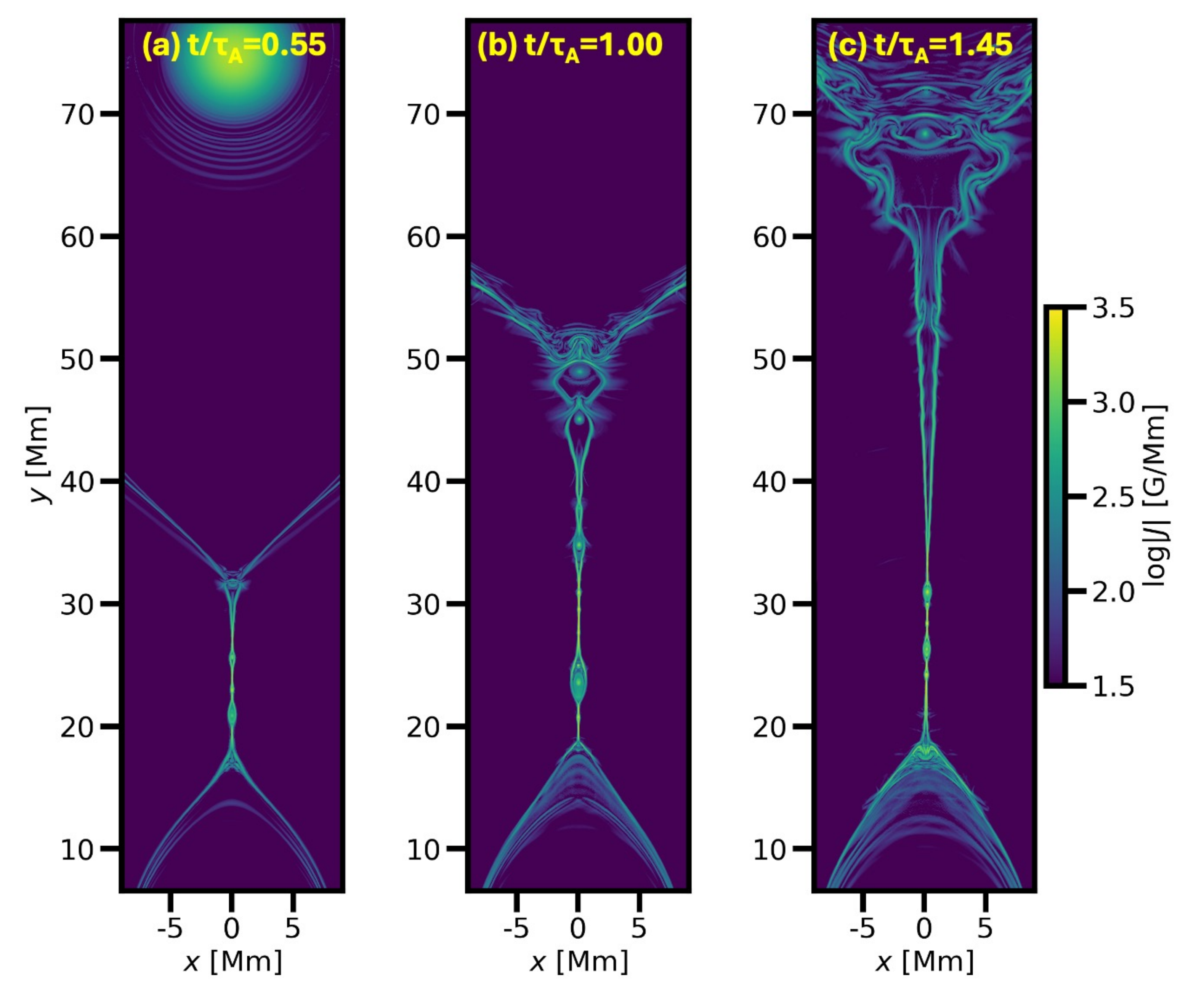}
\caption{Snapshots of the electric current density $\log|J|$ at (a) $t/\tau_A = 0.55$, (b) 1.0, and (c) 1.45. Only the [-10, 10]$\times$[6, 75]$\rm{Mm}^2$ domain is shown here. The eruption of the flux rope induces magnetic reconnection along the vertical current sheet and generates multiple plasmoids. The reconnection layer is distinguished by high current density regions. \label{J_evolve}}
\end{figure}

We carry out 2.5D MHD-SDE simulations of the CME/flare eruption model described by \citet{Chen2020}, in which a pre-existing magnetic flux rope erupts along with a vertical reconnection current sheet \citep[][]{Forbes1995ApJ...446..377F, lin2000JGR...105.2375L}. This model initially exhibits a X-point of magnetic reconnection located below the flux rope, where an extending vertical current sheet gradually develops as the flux rope erupts. The upper section of the vertical current sheet originates from the base of the flux rope and narrows as it extends downward, eventually connecting with its lower counterpart above the flare arcade, which broadens at lower heights (see Figure \ref{J_evolve}). 

We start the simulation without a guide field to capture the impulsive phase when the magnetic energy release and particle acceleration are most efficient.

We normalize the simulations by $L_0$ = 300 Mm, $B_0$ = 50 G, and $n_0=2.5\times10^{9} \rm{cm^{-3}}$. These values are  within the reported ranges of typical parameters of solar flares: magnetic field strengths $B\sim10-100$ G \citep{Lin2000,Yang2020,Brooks2021} and thermal electron number densites of $n_T\sim10^{9}{
\rm cm}^{-3}$ \citep{Warren2009,Young2009,DelZanna2023}. The simulation domain spans $-45<x<45~\rm{Mm}$ and $0<y<180~\rm{Mm}$.
The speed is normalized by the characteristic Alfvén speed is $V_{A0}$ = 2181 km/s, and the corresponding normalized timescale is $\tau_A = L_0/V_{A0} \approx 137$ s.
In our model, the plasma beta in the vertical current sheet has a value of approximately $\beta \approx 0.1$, corresponding to a current sheet temperature of $T_{\rm CS}\approx 10^7$ K.
We adopt a uniform resistivity $\eta=10^{-6}$ resulting in a Lundquist number of $S = L_yV_{A0}/\eta = 6\times10^5$. Although the typical solar coronal conditions correspond to $S = 10^{12}-10^{14}$, the reconnection rate becomes independent of $S$ for $S \gtrsim 10^4$ \citep{bhattacharjee2009,Loureiro2016}. Therefore, our case can represent the reconnection rate of a high-Lundquist-number current sheet. The grid numbers are $n_x \times n_y = 2048 \times 4096$, so the grid resolution is $\Delta x = \Delta y \approx0.04$Mm. The simulation resolves the half-thickness of the current sheet with approximately 10 cells, which is sufficient to capture the magnetic islands that form along the sheet.

%We adopt a spatial diffusion coefficient from quasi-linear theory assuming magnetic turbulence has an isotropic, Kolmogorov-like power spectrum. This is given as \citep{giacalone1999},
%\begin{equation}
%\kappa_\parallel(v) = \frac{3v_{NT}^3}{20L_c\Omega^2_0\sigma^2}\csc\left(\frac{3\pi}{5}\right)\left[1+\frac{72}{7}\left(\frac{\Omega_0L_c}{v_{\rm NT}}\right)^{5/3}\right],
%\end{equation}
%where $v_{\rm NT}$, $L_c$, $\Omega_0$, and $\sigma^2$ represent the nonthermal particle speed, turbulence correlation length, particle gyrofrequency, and the normalized wave variance of turbulence, respectively. In these simulations, we adopt $L_c = 140$km and $\sigma^2=\left<\delta B^2\right>/B_0^2=1$ \citep{Huang2016}. The normalization of the spatial diffusion coefficient is given by $\kappa_0=L_0v_A\approx6.5\times10^{18}~\rm{cm}^2~\rm{s}^{-1}$. In this case, $\kappa_\parallel \approx 6.1\times10^{-5}\kappa_0$ for 10 keV electrons. We assume highly anisotropic spatial diffusion, with $\kappa_\perp = 0.01 \kappa_\parallel$, based on test particle simulations by \citet{giacalone1999}. 

We adopt a spatial diffusion coefficient from quasi-linear theory assuming magnetic turbulence has an isotropic, Kolmogorov-like power spectrum. This is given as \citep{giacalone1999},
\begin{equation}
\kappa_\parallel(p_{\rm NT})
=
\frac{3\,v_{\rm NT}^3}
     {20\,L_c\,\Omega^2_{\rm NT}\,\sigma^2}\,
\csc\!\left(\frac{3\pi}{5}\right)
\left[
  1 + \frac{72}{7}
  \left(
    \frac{\Omega_{\rm NT}\,L_c}{v_{\rm NT}}
  \right)^{5/3}
\right],
\label{eq:kappa_general}
\end{equation}
where $L_c = 140$km is the turbulence correlation length and 
$\sigma^2 = \langle \delta B^2\rangle / B_0^2=1$ \citep{Huang2016} is the normalized wave variance of turbulence.
The gyrofrequency is given by
\begin{eqnarray}
\Omega_{\rm NT} &=& \frac{q B_0}{\gamma_{\rm NT}\,m_e c},
\end{eqnarray}
where $q$ and $\gamma_{\rm NT}$ are the electron charge and the Lorentz factor of nonthermal electrons, respectively. The normalization of the spatial diffusion coefficient is given by $\kappa_0=L_0v_A\approx6.5\times10^{18}~\rm{cm}^2~\rm{s}^{-1}$. In this case, $\kappa_\parallel \approx 6.1\times10^{-5}\kappa_0$ for 10 keV electrons. We assume highly anisotropic spatial diffusion, with $\kappa_\perp = 0.01 \kappa_\parallel$, based on test particle simulations by \citet{giacalone1999}. 
%\textcolor{blue}{The adiabatic index associated with a given nonthermal particle distribution is defined as
%\[
%\gamma = 1 + \frac{P_{\rm NT}}{U_{\rm NT}},
%\]
%where $P_{\rm NT}$ denotes the nonthermal pressure introduced above.
%To compute the corresponding internal energy density, we define the kinetic energy of a
%particle with momentum $p_{\rm NT}$ as
%\[
%K(p_{\rm NT}) = (\gamma(p_{\rm NT}) - 1)\, mc^2.
%\]
%The nonthermal internal energy density is then given by
%\begin{eqnarray}
%U_{\rm NT} &=& \int K(p_{\rm NT})\, f(x_{\rm NT}, p_{\rm NT}, t)\, d^3 p_{\rm NT}.
%\end{eqnarray}
%}

\subsection{Injection of nonthermal electrons}
\label{s2.3}
 
For magnetic reconnection, the initial nonthermal energization can come from Fermi reflection, pickup process, and non-ideal electric field, etc \citep{Guo2024}. Not all of them can be included unless fully kinetic simulations can be used. Fully kinetic simulations have shown that a few tens of percent of thermal electrons can be accelerated into a nonthermal distribution in regions where magnetic reconnection occurs, depending on parameters such as the guide field strength \citep{French2023}. Based on this result, in this paper, we introduce a straightforward approach: a constant fraction of thermal electrons is injected into the nonthermal population within the magnetic reconnection region, identified by high current density. This injected population represents suprathermal particles that would be produced by kinetic pre-acceleration mechanisms such as Fermi reflection in low guide field reconnection \citep[e.g.,][]{Zhang2021, French2023, Guo2024}. Our model treats the injection fraction, $R_i$, as a control parameter due to its uncertainty arising from dependence on local plasma conditions, such as plasma beta and guide field strength.

To set a certain density injection of nonthermal electrons, we first add ``seed particles'' throughout the entire domain. In this study, we set the total number of these particles to be $n_{\rm{seed}}\approx10^9$.
These seed particles are initially massless and do not provide any feedback to the MHD fluid. They are simply advected by the background plasma flow. When they are advected into a high current density region, they are tagged as ``nonthermal'' particles, representing the injection process.  In this simulation, the threshold for the high current density region is set to $|J| \geq 10^3~\rm{G/Mm}$,  but the final result is not very sensitive to the threshold within a range of an order of magnitude. We assign the density of the nonthermal particles proportional to the thermal density to achieve a specific density ratio between nonthermal electrons and the background thermal electron density, $R_i=n_{\rm{NT}}/n_T$. For the fiducial case, we assume that $R_i=10\%$ of thermal electrons are pre-accelerated to nonthermal energies. We also simulate $R_i=1\%$ and $R_i=20\%$ injection cases for comparison. 

In PIC simulation, the injection energy of nonthermal electrons, which corresponds to the starting point of the power-law energy spectrum, is approximately $\approx0.2m_p  v_A^2 \approx 10\rm{keV}$ \citep{Zhang2021,Zhang2024ApJ}, where $m_p$ is the proton mass. Accordingly, we set the injection energy to $\varepsilon_{\rm inj}=~$10 keV.

A particle splitting scheme is adopted to increase statistics, and we set it up in a way that their total mass remains conserved. Once tagged as ``nonthermal particles," particles follow Parker's transport equation and provide feedback as described by equations (\ref{meq}) and (\ref{Eeq}).

When the injection of ``nonthermal particle" occurs, a thermal electron is converted into a nonthermal electron. As a result, the plasma density, momentum density, and energy density within each cell are reduced by the amount corresponding to the injected nonthermal particles by adding sink terms, 
\begin{eqnarray}
    S_\rho = \frac{m_e}{V\Delta t}\int s(p_{\rm NT})dp^3_{\rm NT} ,\\
    S_m = \frac{m_e \boldsymbol{v}}{V\Delta t}\int s(p_{\rm NT})dp^3_{\rm NT},\\
    S_E = \frac{m_e v^2}{2V\Delta t}\int s(p_{\rm NT})dp^3_{\rm NT} + \frac{1}{\Delta t} \int E(p_{\rm NT})s(p_{\rm NT})dp^3_{\rm NT},
\end{eqnarray}
on the right side of equation (\ref{cont})-(\ref{Eeq}), where $V$, $\Delta t$, and $m_e$ are the volume of cell, step size of MHD, and mass of the electron, respectively. $s$ is the distribution function of injected particles. Here, we use a monoenergy injection $s(p_{\rm NT}) = n_{\rm inj}\delta(p_{\rm NT} - p_{\rm inj})$, where $n_{\rm inj}$ denotes the number of nonthermal electrons injected within a cell during the time interval $\Delta t$, and $p_{\rm inj}$ represents the momentum of an electron with 10 keV energy. The second term on the right-hand side of $S_E$ represents the energy transfer from the internal energy of the thermal plasma to nonthermal electrons to ensure energy conservation. The integration occurs for all injected nonthermal particles within $\Delta t$ in each cell.

%This model provides constraints on the physical properties of nonthermal electrons, such as number density and energy density, in physical units. This represents an important advancement in modeling nonthermal electrons in solar flares, allowing us to compare these values with observations.

In our previous study \citep{seo2024feedback}, we assumed a population of pre-existing seed particles as ``pre-selected” to allow particle acceleration through Parker transport equation. A similar approach is also used in the $\it{kglobal}$ simulations \citep[e.g.,][]{Arnold2021}. In this paper, instead of assigning ``pre-selected” particles in the coronal background, we mimic the injection process observed in PIC simulations by injecting ``pre-accelerated” particles at the magnetic reconnection region. We have compared the two models and found quantitatively similar results, which indicate that the choice of initialization of particles is not critical for the final outcome.

%Figure 1

%Figure 1
\begin{figure}[ht!]
\centering
\includegraphics[width=0.8\textwidth]{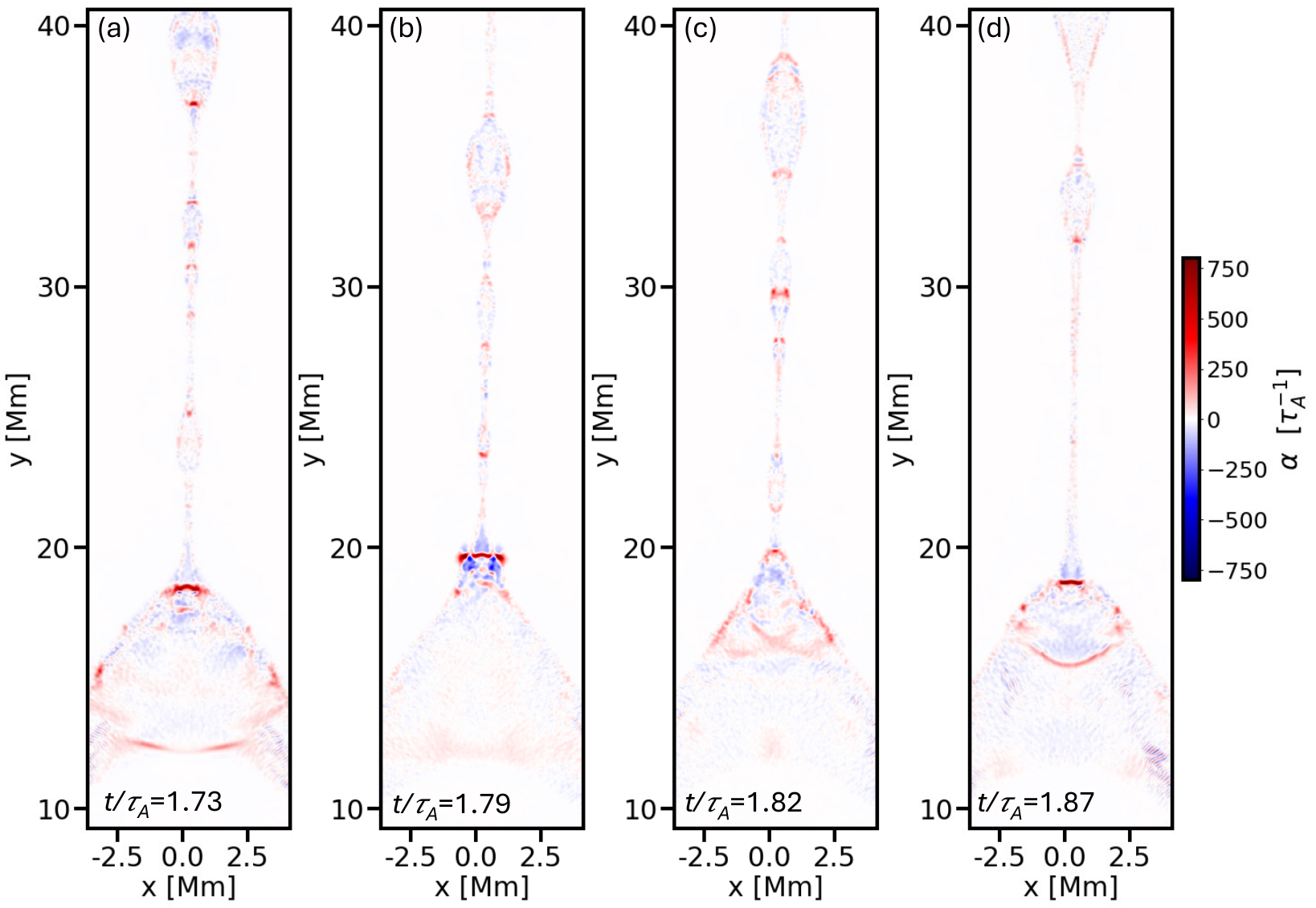}
\caption{Snapshots of the acceleration rate $\alpha$ at (a) $t/\tau_A = 1.73$, (b) 1.79, (c) 1.82, and (d) 1.87 for the $R_i = 10\%$ case. Only the region spanning $[-4, 4] \times [9, 41]~\mathrm{Mm}^2$ is shown. The positive and negative values of $\alpha$ indicate particle acceleration and deceleration, respectively. \label{alpha}}
\end{figure}

\section{Simulation Results}
\label{s3}

\subsection{Evolution of flare eruption model}
The large-scale evolution of the flare eruption model includes the formation of the current sheet and the eruption of the flux rope. We find that feedback from nonthermal electrons does not significantly alter the global characteristics of the eruptive flare, such as the flux rope eruption and magnetic reconnection in the vertical current sheet. While there are some differences in the fine-scale features, such as plasmoid formation and motion, the overall outflow and the morphology of the eruption and looptop remain similar.

The flare eruption model initially shows that the rising flux rope stretches the underlying field lines, forming an elongated current sheet. Reconnection along this vertical current sheet enables the flux rope to successfully erupt from the solar surface \citep[][]{lin2000JGR...105.2375L}. Figure \ref{J_evolve} shows the time evolution of the current density in this model.
As shown in this figure, due to the high $S$, the unstable current sheet narrows to a smaller scale, leading to the formation and rapid growth of a series of magnetic islands, some of which merge into larger structures \citep{Loureiro2007, bhattacharjee2009,Uzdensky2010, Loureiro2016}. The downflow generated by magnetic reconnection impinges on the looptop and creates a termination shock \citep{Shen_2018,kong2019}. These current sheet regions and the termination shock are expected to accelerate particles efficiently.

\subsection{Acceleration of nonthermal electrons}
\label{s3.2}
To identify regions of efficient acceleration during the flare eruption, we track the acceleration rate, $\alpha=(\Delta \varepsilon/\varepsilon)/\Delta t$, where $\varepsilon$ is the kinetic energy of a nonthermal electron. $\alpha$ measures how rapidly particles can be accelerated. In this paper, $\alpha$ is calculated by averaging the acceleration rate of each nonthermal particle, within each cell. $\Delta t$ is one MHD time step, approximately $\approx10^{-3}\tau_A$. 

The momentum evolution in Ito-type stochastic differential equations (SDE) can be described as \citep{li2018},
\begin{equation}
\Delta p_{\rm NT} = -\frac{p_{\rm NT}}{3} (\nabla \cdot \boldsymbol{v}) \Delta t.
\end{equation}
Therefore,
\begin{equation}
\alpha= \frac{\Delta \varepsilon/\varepsilon}{\Delta t} \propto \frac{\Delta p_{\rm NT}/p_{\rm NT}}{\Delta t} \propto  -(\nabla \cdot \boldsymbol{v}).
\end{equation}
Hence, the acceleration efficiency $\alpha$ is proportional to the plasma compression. A positive (negative) value of $\alpha$ corresponds to plasma compression (divergence) and indicates particle acceleration (deceleration).

In Figure \ref{alpha}, we present the evolution of $\alpha$ for $R_i = 10\%$ case. We confirmed that the current sheets and the termination shock are regions that host strong compression for accelerating these electrons. Along the vertical current sheet, magnetic reconnection–driven structures, such as reconnection exhausts and magnetic island mergers, lead to the formation of strong compression regions \citep{li2018, seo2024feedback}. These compression regions have a $\alpha\approx100-300\tau_A^{-1}$. For a detailed discussion of particle acceleration in the magnetic reconnection region using MHD-SDE method, we refer interested readers to \citet{seo2024feedback}. In panel (a), around $y\approx19$ Mm, a fast-mode termination shock is clearly visible in the region with $\alpha\approx600\tau_A^{-1}$. The higher acceleration rate, $\alpha$, observed in this region, compared with that in the flare current sheet, can be explained by the stronger plasma compression.
When large plasmoids (located between $22 \lesssim y \lesssim 27$ Mm in panel (a) of Figure \ref{alpha}) collide with the above-the-looptop region, the acceleration rate at the termination shock is enhanced, as shown in panel (b) around $y \approx 20$ Mm, which can be attributed to stronger plasma compression. Due to this strong compression, the particle acceleration rate at the termination shock is slightly enhanced to $\alpha \approx 700 \tau_A^{-1}$. However, within a few seconds after the initial contact, the collision disrupts the horizontal termination shock structure (panel (c)), consistent with the behavior reported by \citet{Kong2020}.
 The shock is then restored after the plasmoid merges (panel (d)). This phenomenon shows consistency with features observed in X-ray and microwave emissions \citep{Chen2015}.

In this case, the feedback from nonthermal electrons suppresses the compressibility of highly compressed structures, yielding a maximum value of $\alpha$ that is about 60\% lower than in the case without feedback at a similar stage of the flare evolution.

%\textcolor{blue}{As shown in Figure~\ref{alpha} (b)-(d), our model exhibits plasmoid coalescence along the vertical current sheet, which generates regions of enhanced particle acceleration rate and thus indicates rapid electron acceleration. This behavior is consistent with suggestions that plasmoid coalescence can be associated with impulsive emissions in hard X-rays and microwaves \citep{Takasao2016}.}

%Figure 2
\begin{figure}[ht!]
\centering
\includegraphics[width=\textwidth]{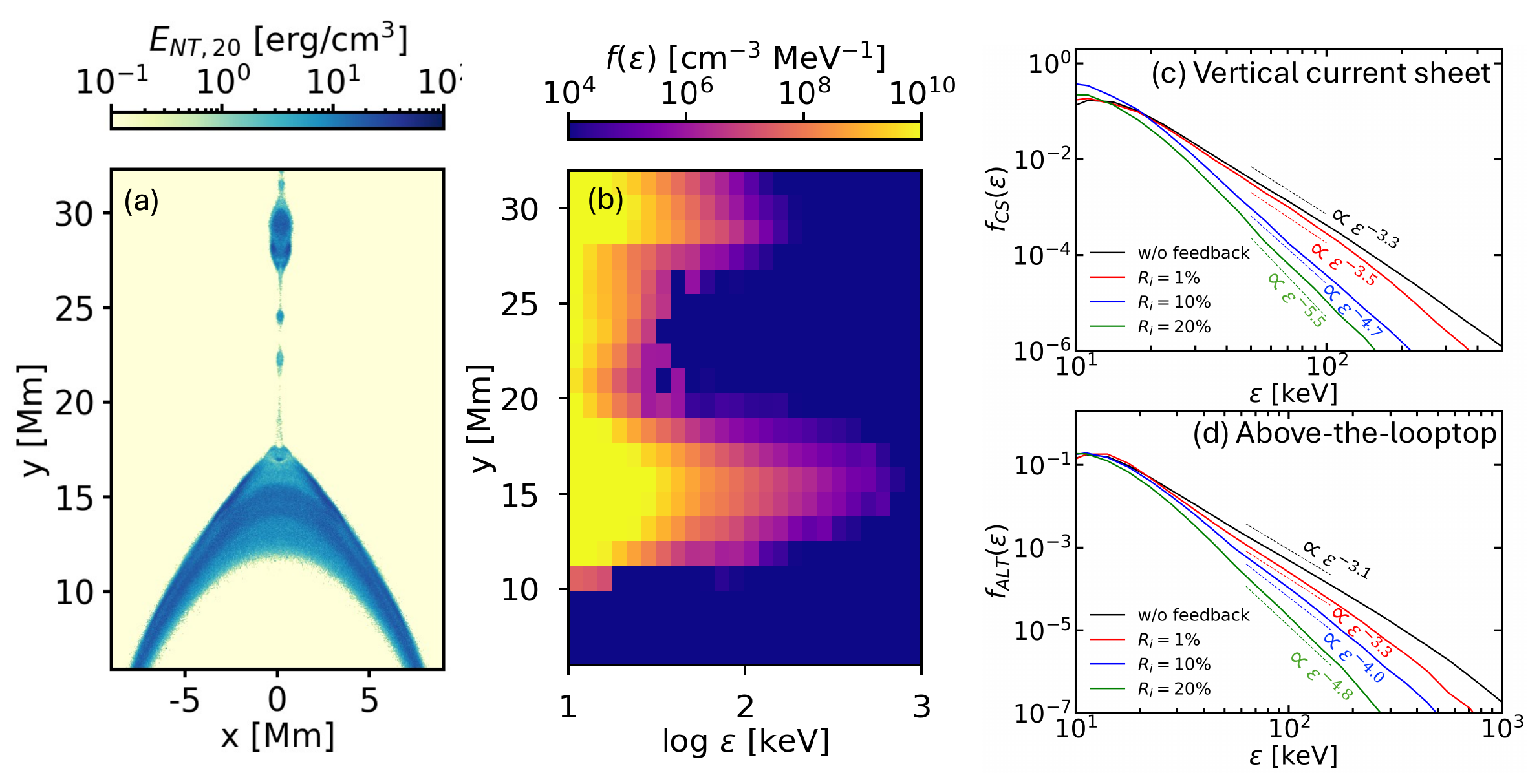}
\caption{(a) The spatial distribution of nonthermal electron energy density with energies above 20 keV. (b) Height–energy distribution of the nonthermal electron along the $y$-direction within the $-2\le x \le 2$ Mm region. (a) and (b) panels are shown at $t/\tau_A = 1.45$ and only in the domain spanning [-10,10]$\times$[6,40]$\rm{Mm}^2$. (c) Energy spectra of nonthermal electrons in the vertical current sheet ($20 \le y \le 40$ Mm), normalized by the total number of particles, for injection rates of 20\% (green solid), 10\% (blue solid), and 1\% (red solid) at $t/\tau_A = 1.45$. Dashed lines indicate the power-law slopes. (d) Same as (c), but for the above-the-looptop region ($6 \le y \le 18$ Mm). \label{E_spec}}
\end{figure}

\subsection{Global distributions of nonthermal electrons and its feedback}

For the first time, our simulation provides the realistic estimate on the energy carried by nonthermal electrons by including their feedback.
The nonthermal electrons are primarily injected around the primary reconnection X-point, $y \sim 21$ Mm in Figure \ref{E_spec} (a) \& (b). Interestingly, this region is characterized by a lack of high-energy electrons because the bidirectional exhaust from the X-point produces flow divergence, and thus compression-driven acceleration does not occur there. Once nonthermal electrons are injected, they are accelerated into high energies due to flow compression in the current sheet region and advected by the reconnection outflows. The nonthermal electrons in the downward outflows are further accelerated by the termination shock to a few hundred keV in the above-the-looptop region ($6 \le  y \le$ 18 Mm in Figure~\ref{E_spec} (a) \& (b)). Figure \ref{E_spec} (a) shows that at the above-the-looptop, nonthermal electrons exceeding 20 keV contain a substantial amount of energy, $E_{\rm{NT},20}=\int^\infty_{20\rm{keV}}\varepsilon f(\varepsilon)d\varepsilon \sim 10~\rm{erg/cm^3}$. For comparison, in the case with $R_i=1\%$, $E_{\rm{NT},20}$ is on the order of unity, whereas in the case with $R_i=20\%$, it has a similar value to the $R_i=10\%$ case (see Figure \ref{e_evolve}).
 Due to the additional acceleration at the termination shock, most of these high-energy electrons are concentrated in the above-the-looptop region, with a density two orders of magnitude higher than that in the current sheet. These trends are also found in the EOVSA observations \citep{Chen2020}. They showed a lack of high-energy nonthermal particles with energies above 100 keV at the primary reconnection X-point and an enrichment at the above-the-looptop region.

To illustrate the role of feedback, we compare the energy spectrum in different runs and show the results in Figure~\ref{E_spec} (c) \& (d). In all these runs, the nonthermal electrons develop power-law energy spectra, and the spectral indices are within the range often observed \citep{kruker2010,oka2013looptop,oka2015,Effenberger2017}. For example, in the fiducial case ($R_i=10\%$), the spectrum in the vertical current sheet $f_{\rm{CS}}(\varepsilon)\propto\varepsilon^{-4.7}$, and the spectrum in the above-the-looptop region $f_{\rm{ATL}}(\varepsilon)\propto\varepsilon^{-4.0}$, which is harder than $f_{\rm{CS}}$ due to the additional acceleration. 

The nonthermal electron pressure provides feedback to the thermal plasma by suppressing compression structures, which in turn mediates particle energization. As a result, greater feedback from energetic particles leads to a steeper power-law energy spectrum \citep{seo2024feedback}.
So, as shown in Figure~\ref{E_spec} (c) \& (d), if we increase (decrease) the nonthermal fraction $R_i$, the spectrum will become steeper (harder).  
Thus, the feedback from nonthermal electrons is one of the key factors in electron acceleration and in determining the slope of the energy spectrum in solar flares. We note that, in Fermi-type acceleration mechanisms, the spectral index is primarily governed by the acceleration rate and the particle escape timescale \citep[e.g.,][]{Drury1983,Guo2014,li2019power-law}. This underscores that particle escape, in addition to acceleration, plays a fundamental role in determining the resulting energy spectrum.

%Figure 3
\begin{figure}[ht!]
\centering
\includegraphics[width=0.7\textwidth]{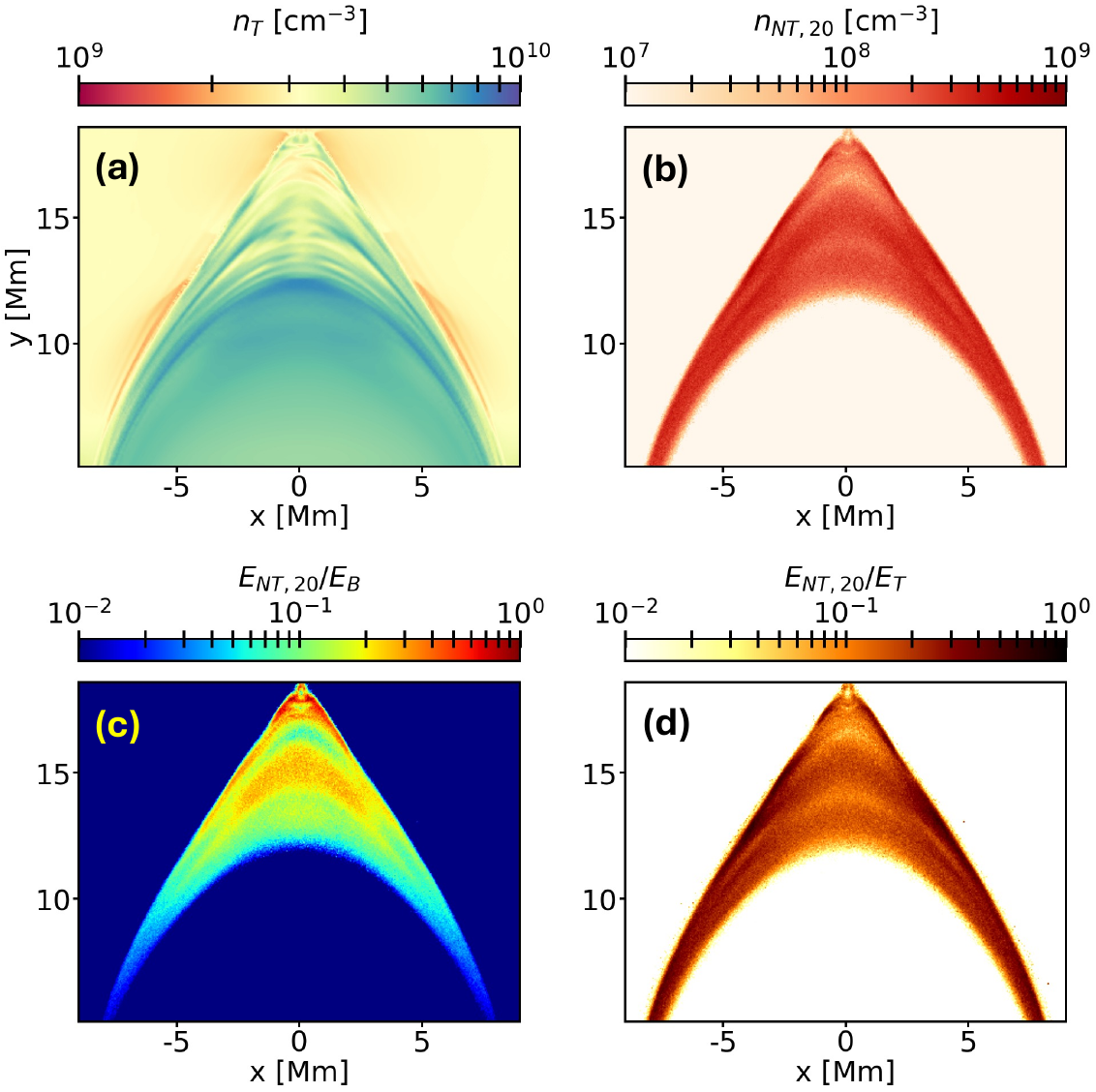}
\caption{A snapshot showing (a) the thermal electron density $n_T$, (b) the nonthermal electron density, $n_{\rm{NT}}$, (c) the ratio of nonthermal electron energy to magnetic energy, $E_{\rm{NT}}/E_B$, and (d) the ratio between nonthermal electron energy to thermal electron energy at $t/\tau_A = 1.45$. Only the [-10,10]$\times$[6,19]$\rm{Mm}^2$ domain is shown here.\label{Spatial}}
\end{figure}

\subsection{Energetic electrons distribution in the above-the-looptop region}

In this subsection, we focus on the features found in the above-the-looptop region, where nonthermal electrons are predominantly present.
We plot (a) the thermal electron density, $n_T$, (b) nonthermal electron number density, $n_{\rm NT}$, (c) the ratio of nonthermal electron energy density to magnetic energy density $E_{\rm NT,20}/E_B$, and (d) the ratio of nonthermal electron energy density to thermal plasma energy density $E_{\rm NT,20}/E_T$ in Figure~\ref{Spatial} for $R_i=10\%$ case. Panel (a) reveals that the thermal electron density in the above-the-looptop region is several times lower than that in the magnetic loop region. This low-density region forms because the plasma is supplied by magnetic reconnection occurring higher up in the corona, where the background density is much lower. In this region, thermal electron number density is $n_T\approx3-6\times10^{9}~\mathrm{cm^{-3}}$. This value lies within the range observed for the thermal electron density in this region, which is $10^{9}–10^{10}~\mathrm{cm}^{-3}$ \citep{krucker2010, oka2013looptop, oka2015, kontar2023}. Panel (b) shows that energetic particles are primarily concentrated in the above-the-looptop region, where $n_{\rm{NT},20}=\int^\infty_{20\rm{keV}} f(\varepsilon)d\varepsilon$. In this region, the magnetic field exhibits a local minimum resembling a “magnetic bottle” \citep{Chen2020}, and as a result the ratio $E_{\rm NT,20}/E_B$ can reach several tens of percent, as illustrated in panel (c). Moreover, the energy density of nonthermal electrons can amount to nearly half of the thermal plasma energy, $E_{\rm NT,20}/E_T \lesssim 50\%$, making it comparable in magnitude, as indicated in panel (d). We also observe higher values of $n_{\rm NT,20}/n_T$, $E_{\rm NT,20}/E_B$, and $E_{\rm NT,20}/E_T$ for larger $R_i$, demonstrating the model’s ability to produce dominant energetic nonthermal electrons in the above-the-looptop region and further emphasizing the importance of including the feedback from energetic electrons.

%Figure 3
\begin{figure}[ht!]
\centering
\includegraphics[width=0.7\textwidth]{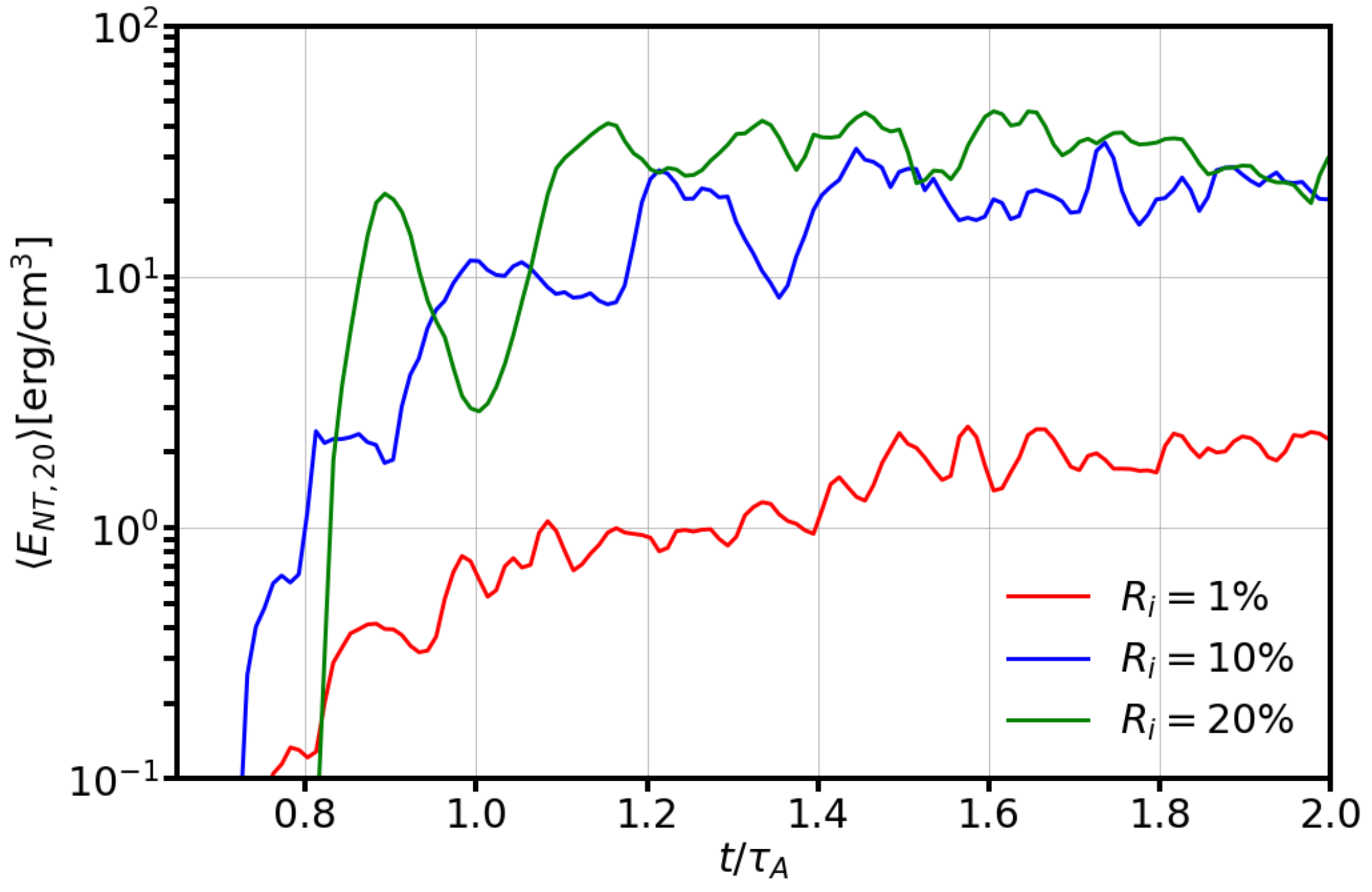}
\caption{The time evolution of the mean energy density of nonthermal electrons with energies above 20 keV within the above-the-looptop region ($6\le y \le 18\rm{Mm}$ with nonthermal electron presence). The period of oscillation is $\sim 0.1\tau_A = $ a few 10s of seconds, which is similar to the period of QPPs \citep{hayes2020}.\label{e_evolve}}
\end{figure}

The nonthermal electrons also exhibit significant temporal variability, as illustrated in Figure \ref{e_evolve}, which shows the time evolution of the nonthermal electron energy density with energy above 20 keV at the above-the-looptop region. The energy density increases up to $t/\tau_A \approx 1$ as the evolution of the current sheet supplies nonthermal electrons to the above-the-looptop region, and the termination shock effectively accelerates them. After $t/\tau_A \approx 1$, the acceleration and escape of nonthermal electrons become nearly balanced, resulting in a saturation in the mean energy density. 
In this quasi–steady state, particle escape plays an important role in determining the energy density in this region.
During the saturation period, the mean energy density of nonthermal electrons with energies above 20 keV in the above-the-looptop region is approximately $10$–$30~\mathrm{erg/cm^3}$ for the $R_i = 10\%$ and $20\%$ cases. On top of this evolution, oscillations are observed, which correspond to the periodic impacts of large magnetic islands on the above-the-looptop region. Hence, variations in the energy density of nonthermal electrons may give rise to quasi-periodic pulsations (QPPs) observed in solar flares across multiple wavelengths \citep{kane1983, inglis2008, hayes2020}. The period of this variation, $\sim 0.1\tau_A \approx$ a few tens of seconds, lies within the observed range of QPP periods. A similar phenomenon was observed during the 2010 August 18 solar flare, where plasmoid collisions with flare loops coincided with impulsive microwave bursts at 34 GHz attributed to gyrosynchrotron emission from nonthermal electrons \citep{Takasao2016}. In addition, \citet{asai2004} reported that downward-moving plasmoids show a strong temporal correlation with impulsive nonthermal emissions.

%Figure 4
\begin{figure}[ht!]
\centering
\includegraphics[width=\textwidth]{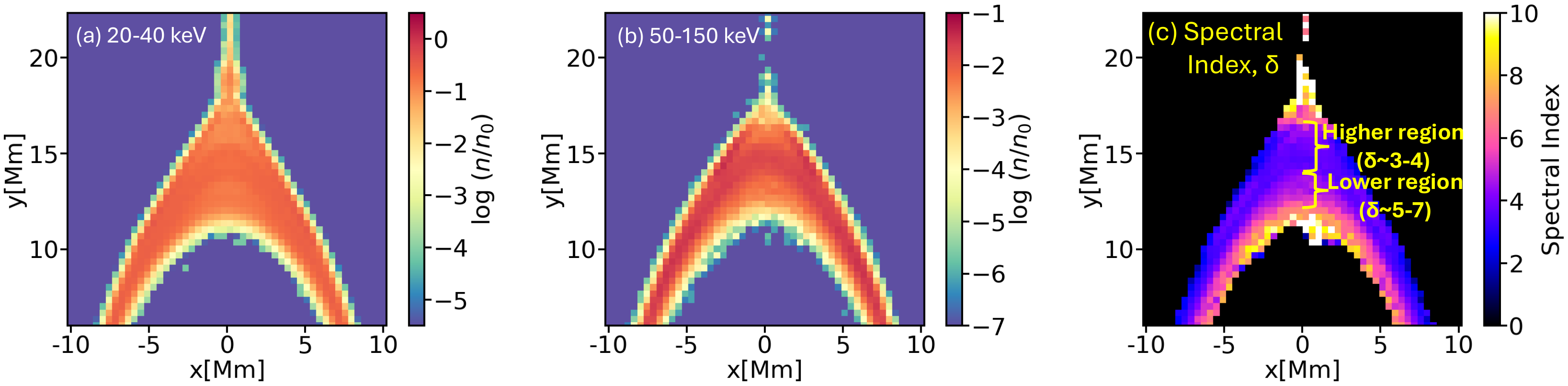}
\caption{(a) Spatial density distribution of nonthermal electrons in the 20–40 keV range.
(b) Spatial density distribution of nonthermal electrons in the 50–150 keV range. (c) The spectral index of nonthermal electrons around the flare looptop region at $t/\tau_A = 1.45$. The spectral index is measured in the range of 30 keV to 100 keV. Only the [-10,10] $\times$ [6,22]$~\rm{Mm^2}$ domain is shown here. \label{S_index}}
\end{figure}

Nonthermal electron distribution at the above-the-looptop region is also energy dependent.
Figure \ref{S_index} (a) \& (b) show the nonthermal electron density distribution for the 20–40 keV and 50–150 keV, respectively, at $t/\tau_A=1.45$. The nonthermal electrons are advected by the reconnection outflows and accumulate in the above-the-looptop region. Near the termination shock, these electrons undergo Fermi acceleration and are further energized, producing a higher-energy nonthermal population. Hence, as shown in Figure \ref{S_index}(a) and (b), compared to 20–40 keV electrons, 50–150 keV electrons are more prevalent in the higher region ($y > 13$ Mm) than in the lower region ($y < 13$ Mm). This trend is consistent with the results of \citet{kong2019}, which showed that higher-energy electrons are concentrated closer to the termination shock at higher altitudes, due to acceleration by the shock and confinement within a concave-downward magnetic trap.
As a result, the electron spectrum is harder in the higher region than in the lower region, as shown in Figure \ref{S_index} (c). The power-law spectral index $\delta\sim3-4$ in the higher above-the-looptop region, while $\delta\sim5-7$ in the lower region. 

In the above-the-looptop region, the power-law energy spectrum arises from first-order Fermi acceleration driven by strong plasma compression at the magnetic reconnection site and the termination shock \citep{kong2019,li2022,kong2022,Zhang2024ApJ}. The spectral slope is largely determined by the local compression ratio. While thermal conduction in hot plasma can alter the thermodynamic structure and even modify the geometry of termination shocks \citep{Takasao2016b,Shen_2018}. 
In addition, when thermal conduction is included, the reconnection outflow Mach number depends on plasma beta \citep{Yokoyama1998, Yokoyama2001, Seaton2009, Takasao2016b}. Notably, as shown in Figure 9 of \citet{Seaton2009}, the effect of thermal conduction becomes significant for $\beta \lesssim 0.01$.
As discussed earlier, our current model does not incorporate several physical effects, such as thermal conduction, pressure anisotropy, and magnetic mirroring. Nevertheless, the model remains physically meaningful because it captures the essential role of feedback from nonthermal electrons in regulating particle acceleration within the flare geometry. A more comprehensive investigation that incorporates these additional multiphysics processes is deferred to future work.

%\textcolor{blue}{\citet{Shen_2018} showed that, under the coronal condition that $\beta=0.1$ and $n_0\sim10^{10}\rm{cm^{-3}}$, thermal conduction does not strongly modify the shock compression ratio. However, in lower-density  conditions, thermal conduction can be a stronger \citep{Yokoyama2001}, influence on the mach number of reconnection outflow and may alter the effective compression ratio of the termination shock \citep{Takasao2016b}. Therefore, while our use of adiabatic compression in the Parker transport equation provides a physically reasonable approximation for the present parameter regime, the resulting compression ratio, and thus the spectral index, may differ in environments where thermal conduction plays a significant role. We defer a study with more comprehensive multiphysics effects in the future.}

\section{Summary}
\label{s4}
For the first time, we perform simulations of particle acceleration and transport in a flare eruption model, incorporating the feedback of nonthermal electrons into the MHD plasma. We find a prominent presence of high-energy nonthermal electrons in the above-the-looptop region, resulting from acceleration by the termination shock and magnetic reconnection. The main findings are summarized as follows,

1. In the fiducial model, where the injection rate is $R_i=n_{\rm NT}/n_T=10\%$, we find a power-law energy spectrum for nonthermal electrons at the above-the-looptop region, $f_{\rm{ATL}}(\varepsilon) \propto \varepsilon^{-4}$, with the slope depending on the injection rate, $R_i$, of the nonthermal electrons. 

2. As effective acceleration occurs, nonthermal electrons at the above-the-looptop region are accelerated to energies above $\sim$100 keV, which can serve as a potential explanation of observed hard X-rays and microwaves \citep{kruker2010,oka2013looptop,oka2015,Effenberger2017,Chen2020,kontar2023}.

3. At the above-the-looptop region, the energy carried by nonthermal electrons corresponds to roughly several tens of percent of the magnetic energy, $E_{\rm NT,20}/E_B \approx 10-30\%$, and up to about half of the thermal energy, $E_{\rm NT,20}/E_T \lesssim 50\%$, underscoring the importance of including feedback from nonthermal electrons.

4. Our model yields an energy density of nonthermal electrons with energies above 20 keV of $E_{\rm{NT},20} \approx 10–30~\mathrm{erg~cm^{-3}}$ in the above-the-looptop region. The impact of large magnetic islands on the above-the-looptop region occurs with a period of several tens of seconds, leading to oscillations in the energy of nonthermal electrons. These oscillations may result in the generation of quasi-periodic pulsations \citep{kane1983, inglis2008, hayes2020}.

In conclusion, our study on the acceleration of nonthermal electrons in the flare model including above-the-looptop region and show that these have a power-law energy spectrum, which is a potential explanation for coronal emissions during solar flares, even when their feedback is taken into account.

\section*{Acknowledgements}
We acknowledge the support from Los Alamos National Laboratory through the LDRD program, DOE OFES, and NASA programs through grant NNH17AE68I,
80HQTR20T0073, and 80HQTR21T0005, 80HQTR21T0117, NNH230B17A, and NNH240B72A. J.S. acknowledges the support from the Los Alamos National Laboratory (LANL) through its Center for Space and Earth Science (CSES). CSES is funded by LANL’s Laboratory Directed Research and Development (LDRD)program under project number 20240477CR. F. G. acknowledges NSF Award No. 2334932. X.L. acknowledges the support from NASA grant 80NSSC21K1313, National Science Foundation grant No. AST-2107745 and AGS-2334930, University of California, Berkeley through subcontract No. 00011753, and Smithsonian Astrophysical Observatory through subcontract No. SV1-21012. B.C. acknowledges the support from NSF grant AST-2108853 to New Jersey Institute of Technology (NJIT) and NASA grant NNH240B72A (through subcontract 89233218CNA000001/C5509 from LANL to NJIT).
The simulations used resources provided by the Los Alamos National Laboratory Institutional Computing
Program, which is supported by the U.S. Department of Energy National Nuclear Security Administration under Contract No. 89233218CNA000001, the National Energy Research Scientific Computing Center (NERSC), a Department of Energy Office of Science User Facility using NERSC award FES-ERCAP0031616, and the Texas Advanced Computing Center (TACC). 

\bibliography{refs}{}

@ARTICLE{lin2000JGR...105.2375L,
       author = {{Lin}, J. and {Forbes}, T.~G.},
        title = "{Effects of reconnection on the coronal mass ejection process}",
      journal = {\jgr},
         year = 2000,
        month = feb,
       volume = {105},
       number = {A2},
        pages = {2375-2392},
          doi = {10.1029/1999JA900477},
       adsurl = {https://ui.adsabs.harvard.edu/abs/2000JGR...105.2375L}
}

@ARTICLE{Forbes1995ApJ...446..377F,
       author = {{Forbes}, T.~G. and {Priest}, E.~R.},
        title = "{Photospheric Magnetic Field Evolution and Eruptive Flares}",
      journal = {\apj},
         year = 1995,
        month = jun,
       volume = {446},
        pages = {377},
          doi = {10.1086/175797},
       adsurl = {https://ui.adsabs.harvard.edu/abs/1995ApJ...446..377F}
}

@ARTICLE{Effenberger2017,
       author = {{Effenberger}, Frederic and {Rubio da Costa}, Fatima and {Oka}, Mitsuo and {Saint-Hilaire}, Pascal and {Liu}, Wei and {Petrosian}, Vah{\'e} and {Glesener}, Lindsay and {Krucker}, S{\"a}m},
        title = "{Hard X-Ray Emission from Partially Occulted Solar Flares: RHESSI Observations in Two Solar Cycles}",
      journal = {\apj},
         year = 2017,
        month = feb,
       volume = {835},
       number = {2},
          eid = {124},
        pages = {124},
          doi = {10.3847/1538-4357/835/2/124},
archivePrefix = {arXiv},
       eprint = {1612.02856},
 primaryClass = {astro-ph.SR},
       adsurl = {https://ui.adsabs.harvard.edu/abs/2017ApJ...835..124E}
}

@ARTICLE{kong2022,
       author = {{Kong}, Xiangliang and {Chen}, Bin and {Guo}, Fan and {Shen}, Chengcai and {Li}, Xiaocan and {Ye}, Jing and {Zhao}, Lulu and {Jiang}, Zelong and {Yu}, Sijie and {Chen}, Yao and {Giacalone}, Joe},
        title = "{Numerical Modeling of Energetic Electron Acceleration, Transport, and Emission in Solar Flares: Connecting Loop-top and Footpoint Hard X-Ray Sources}",
      journal = {\apjl},
         year = 2022,
        month = dec,
       volume = {941},
       number = {2},
          eid = {L22},
        pages = {L22},
          doi = {10.3847/2041-8213/aca65c},
archivePrefix = {arXiv},
       eprint = {2211.15333},
 primaryClass = {astro-ph.SR},
       adsurl = {https://ui.adsabs.harvard.edu/abs/2022ApJ...941L..22K}
}

@ARTICLE{Zhang2021,
       author = {{Zhang}, Qile and {Guo}, Fan and {Daughton}, William and {Li}, Hui and {Li}, Xiaocan},
        title = "{Efficient Nonthermal Ion and Electron Acceleration Enabled by the Flux-Rope Kink Instability in 3D Nonrelativistic Magnetic Reconnection}",
      journal = {\prl},
         year = 2021,
        month = oct,
       volume = {127},
       number = {18},
          eid = {185101},
        pages = {185101},
          doi = {10.1103/PhysRevLett.127.185101},
archivePrefix = {arXiv},
       eprint = {2105.04521},
 primaryClass = {astro-ph.SR},
       adsurl = {https://ui.adsabs.harvard.edu/abs/2021PhRvL.127r5101Z}
}

@article{wang2022,
doi = {10.3847/1538-4357/ac6ddc},
url = {https://dx.doi.org/10.3847/1538-4357/ac6ddc},
year = {2022},
month = {jun},
publisher = {The American Astronomical Society},
volume = {932},
number = {1},
pages = {65},
author = {B.-B. Wang and G. P. Zank and L.-L. Zhao and L. Adhikari},
title = {Turbulent Cosmic Ray–Mediated Shocks in the Hot Ionized Interstellar Medium},
journal = {The Astrophysical Journal}
}

@ARTICLE{drake2019,
       author = {{Drake}, J.~F. and {Arnold}, H. and {Swisdak}, M. and {Dahlin}, J.~T.},
        title = "{A computational model for exploring particle acceleration during reconnection in macroscale systems}",
      journal = {Physics of Plasmas},
         year = 2019,
        month = jan,
       volume = {26},
       number = {1},
          eid = {012901},
        pages = {012901},
          doi = {10.1063/1.5058140},
archivePrefix = {arXiv},
       eprint = {1809.04568},
 primaryClass = {astro-ph.SR},
       adsurl = {https://ui.adsabs.harvard.edu/abs/2019PhPl...26a2901D}
}

@ARTICLE{bai2015,
       author = {{Bai}, Xue-Ning and {Caprioli}, Damiano and {Sironi}, Lorenzo and {Spitkovsky}, Anatoly},
        title = "{Magnetohydrodynamic-particle-in-cell Method for Coupling Cosmic Rays with a Thermal Plasma: Application to Non-relativistic Shocks}",
      journal = {\apj},
         year = 2015,
        month = aug,
       volume = {809},
       number = {1},
          eid = {55},
        pages = {55},
          doi = {10.1088/0004-637X/809/1/55},
archivePrefix = {arXiv},
       eprint = {1412.1087},
 primaryClass = {astro-ph.HE},
       adsurl = {https://ui.adsabs.harvard.edu/abs/2015ApJ...809...55B}
}

@ARTICLE{li2018,
       author = {{Li}, Xiaocan and {Guo}, Fan and {Li}, Hui and {Li}, Shengtai},
        title = "{Large-scale Compression Acceleration during Magnetic Reconnection in a Low-{\ensuremath{\beta}} Plasma}",
      journal = {\apj},
         year = 2018,
        month = oct,
       volume = {866},
       number = {1},
          eid = {4},
        pages = {4},
          doi = {10.3847/1538-4357/aae07b},
archivePrefix = {arXiv},
       eprint = {1807.03427},
 primaryClass = {astro-ph.SR},
       adsurl = {https://ui.adsabs.harvard.edu/abs/2018ApJ...866....4L}
}

@ARTICLE{giacalone1999,
       author = {{Giacalone}, J. and {Jokipii}, J.~R.},
        title = "{The Transport of Cosmic Rays across a Turbulent Magnetic Field}",
      journal = {\apj},
         year = 1999,
        month = jul,
       volume = {520},
       number = {1},
        pages = {204-214},
          doi = {10.1086/307452},
       adsurl = {https://ui.adsabs.harvard.edu/abs/1999ApJ...520..204G}
}

@ARTICLE{li2022,
       author = {{Li}, Xiaocan and {Guo}, Fan and {Chen}, Bin and {Shen}, Chengcai and {Glesener}, Lindsay},
        title = "{Modeling Electron Acceleration and Transport in the Early Impulsive Phase of the 2017 September 10th Solar Flare}",
      journal = {\apj},
         year = 2022,
        month = jun,
       volume = {932},
       number = {2},
          eid = {92},
        pages = {92},
          doi = {10.3847/1538-4357/ac6efe},
archivePrefix = {arXiv},
       eprint = {2205.04946},
 primaryClass = {astro-ph.SR},
       adsurl = {https://ui.adsabs.harvard.edu/abs/2022ApJ...932...92L}
}

@ARTICLE{kruker2010,
       author = {{Krucker}, S{\"a}m and {Hudson}, H.~S. and {Glesener}, L. and {White}, S.~M. and {Masuda}, S. and {Wuelser}, J. -P. and {Lin}, R.~P.},
        title = "{Measurements of the Coronal Acceleration Region of a Solar Flare}",
      journal = {\apj},
         year = 2010,
        month = may,
       volume = {714},
       number = {2},
        pages = {1108-1119},
          doi = {10.1088/0004-637X/714/2/1108},
       adsurl = {https://ui.adsabs.harvard.edu/abs/2010ApJ...714.1108K}
}

@ARTICLE{parker1965,
       author = {{Parker}, E.~N.},
        title = "{The passage of energetic charged particles through interplanetary space}",
      journal = {\planss},
         year = 1965,
        month = jan,
       volume = {13},
       number = {1},
        pages = {9-49},
          doi = {10.1016/0032-0633(65)90131-5},
       adsurl = {https://ui.adsabs.harvard.edu/abs/1965P&SS...13....9P}
}

@ARTICLE{guo2014,
       author = {{Guo}, Fan and {Li}, Hui and {Daughton}, William and {Liu}, Yi-Hsin},
        title = "{Formation of Hard Power Laws in the Energetic Particle Spectra Resulting from Relativistic Magnetic Reconnection}",
      journal = {\prl},
         year = 2014,
        month = oct,
       volume = {113},
       number = {15},
          eid = {155005},
        pages = {155005},
          doi = {10.1103/PhysRevLett.113.155005},
archivePrefix = {arXiv},
       eprint = {1405.4040},
 primaryClass = {astro-ph.HE},
       adsurl = {https://ui.adsabs.harvard.edu/abs/2014PhRvL.113o5005G}
}

@ARTICLE{kong2019,
       author = {{Kong}, Xiangliang and {Guo}, Fan and {Shen}, Chengcai and {Chen}, Bin and {Chen}, Yao and {Musset}, Sophie and {Glesener}, Lindsay and {Pongkitiwanichakul}, Peera and {Giacalone}, Joe},
        title = "{The Acceleration and Confinement of Energetic Electrons by a Termination Shock in a Magnetic Trap: An Explanation for Nonthermal Loop-top Sources during Solar Flares}",
      journal = {\apjl},
         year = 2019,
        month = dec,
       volume = {887},
       number = {2},
          eid = {L37},
        pages = {L37},
          doi = {10.3847/2041-8213/ab5f67},
archivePrefix = {arXiv},
       eprint = {1911.08064},
 primaryClass = {astro-ph.SR},
       adsurl = {https://ui.adsabs.harvard.edu/abs/2019ApJ...887L..37K}
}

@article{arnold2021,
  title = {Electron Acceleration during Macroscale Magnetic Reconnection},
  author = {Arnold, H. and Drake, J. F. and Swisdak, M. and Guo, F. and Dahlin, J. T. and Chen, B. and Fleishman, G. and Glesener, L. and Kontar, E. and Phan, T. and Shen, C.},
  journal = {Phys. Rev. Lett.},
  volume = {126},
  issue = {13},
  pages = {135101},
  numpages = {7},
  year = {2021},
  month = {Mar},
  publisher = {American Physical Society},
  doi = {10.1103/PhysRevLett.126.135101},
  url = {https://link.aps.org/doi/10.1103/PhysRevLett.126.135101}
}

@ARTICLE{sun2023,
       author = {{Sun}, Xiaochen and {Bai}, Xue-Ning},
        title = "{The magnetohydrodynamic-particle-in-cell module in ATHENA++: implementation and code tests}",
      journal = {\mnras},
         year = 2023,
        month = aug,
       volume = {523},
       number = {3},
        pages = {3328-3347},
          doi = {10.1093/mnras/stad1548},
archivePrefix = {arXiv},
       eprint = {2304.10568},
 primaryClass = {astro-ph.HE},
       adsurl = {https://ui.adsabs.harvard.edu/abs/2023MNRAS.523.3328S}
}

@ARTICLE{ruszkowski2023,
       author = {{Ruszkowski}, Mateusz and {Pfrommer}, Christoph},
        title = "{Cosmic ray feedback in galaxies and galaxy clusters}",
      journal = {\aapr},
         year = 2023,
        month = dec,
       volume = {31},
       number = {1},
          eid = {4},
        pages = {4},
          doi = {10.1007/s00159-023-00149-2},
archivePrefix = {arXiv},
       eprint = {2306.03141},
 primaryClass = {astro-ph.HE},
       adsurl = {https://ui.adsabs.harvard.edu/abs/2023A&ARv..31....4R}
}

@ARTICLE{loureiro2007,
       author = {{Loureiro}, N.~F. and {Schekochihin}, A.~A. and {Cowley}, S.~C.},
        title = "{Instability of current sheets and formation of plasmoid chains}",
      journal = {Physics of Plasmas},
         year = 2007,
        month = oct,
       volume = {14},
       number = {10},
        pages = {100703-100703},
          doi = {10.1063/1.2783986},
archivePrefix = {arXiv},
       eprint = {astro-ph/0703631},
 primaryClass = {astro-ph},
       adsurl = {https://ui.adsabs.harvard.edu/abs/2007PhPl...14j0703L}
}

@ARTICLE{bhattacharjee2009,
       author = {{Bhattacharjee}, A. and {Huang}, Yi-Min and {Yang}, H. and {Rogers}, B.},
        title = "{Fast reconnection in high-Lundquist-number plasmas due to the plasmoid Instability}",
      journal = {Physics of Plasmas},
         year = 2009,
        month = nov,
       volume = {16},
       number = {11},
          eid = {112102},
        pages = {112102},
          doi = {10.1063/1.3264103},
archivePrefix = {arXiv},
       eprint = {0906.5599},
 primaryClass = {physics.plasm-ph},
       adsurl = {https://ui.adsabs.harvard.edu/abs/2009PhPl...16k2102B}
}

@ARTICLE{gary2018,
   	author = {{Gary}, Dale E. and {Chen}, Bin and {Dennis}, Brian R. and {Fleishman}, Gregory D. and {Hurford}, Gordon J. and {Krucker}, S{\"a}m and {McTiernan}, James M. and {Nita}, Gelu M. and {Shih}, Albert Y. and {White}, Stephen M. and {Yu}, Sijie},
    	title = "{Microwave and Hard X-Ray Observations of the 2017 September 10 Solar Limb Flare}",
  	journal = {\apj},
     	year = 2018,
    	month = aug,
   	volume = {863},
   	number = {1},
      	eid = {83},
    	pages = {83},
      	doi = {10.3847/1538-4357/aad0ef},
archivePrefix = {arXiv},
   	eprint = {1807.02498},
 primaryClass = {astro-ph.SR},
   	adsurl = {https://ui.adsabs.harvard.edu/abs/2018ApJ...863...83G}
}

@ARTICLE{chen2020,
       author = {{Chen}, Bin and {Shen}, Chengcai and {Gary}, Dale E. and {Reeves}, Katharine K. and {Fleishman}, Gregory D. and {Yu}, Sijie and {Guo}, Fan and {Krucker}, S{\"a}m and {Lin}, Jun and {Nita}, Gelu M. and {Kong}, Xiangliang},
        title = "{Measurement of magnetic field and relativistic electrons along a solar flare current sheet}",
      journal = {Nature Astronomy},
         year = 2020,
        month = jan,
       volume = {4},
        pages = {1140-1147},
          doi = {10.1038/s41550-020-1147-7},
archivePrefix = {arXiv},
       eprint = {2005.12757},
 primaryClass = {astro-ph.SR},
       adsurl = {https://ui.adsabs.harvard.edu/abs/2020NatAs...4.1140C}
}

@ARTICLE{krucker2010,
       author = {{Krucker}, S{\"a}m and {Hudson}, H.~S. and {Glesener}, L. and {White}, S.~M. and {Masuda}, S. and {Wuelser}, J. -P. and {Lin}, R.~P.},
        title = "{Measurements of the Coronal Acceleration Region of a Solar Flare}",
      journal = {\apj},
         year = 2010,
        month = may,
       volume = {714},
       number = {2},
        pages = {1108-1119},
          doi = {10.1088/0004-637X/714/2/1108},
       adsurl = {https://ui.adsabs.harvard.edu/abs/2010ApJ...714.1108K}
}

@ARTICLE{oka2015,
       author = {{Oka}, Mitsuo and {Krucker}, S{\"a}m and {Hudson}, Hugh S. and {Saint-Hilaire}, Pascal},
        title = "{Electron Energy Partition in the Above-the-looptop Solar Hard X-Ray Sources}",
      journal = {\apj},
         year = 2015,
        month = feb,
       volume = {799},
       number = {2},
          eid = {129},
        pages = {129},
          doi = {10.1088/0004-637X/799/2/129},
       adsurl = {https://ui.adsabs.harvard.edu/abs/2015ApJ...799..129O}
}

@ARTICLE{carmichael1964,
       author = {{Carmichael}, H.},
        title = "{A Process for Flares}",
    journal = {NASA Special Publication},
         year = 1964,
         volume = {50},
        pages = {451},
       adsurl = {https://ui.adsabs.harvard.edu/\#abs/1964NASSP..50..451C}
}

@ARTICLE{sturrock1966,
       author = {{Sturrock}, P.~A.},
        title = "{Model of the High-Energy Phase of Solar Flares}",
      journal = {\nat},
         year = "1966",
        month = "Aug",
       volume = {211},
       number = {5050},
        pages = {695-697},
          doi = {10.1038/211695a0},
       adsurl = {https://ui.adsabs.harvard.edu/abs/1966Natur.211..695S}
}

@ARTICLE{hirayama1974,
       author = {{Hirayama}, T.},
        title = "{Theoretical Model of Flares and Prominences. I: Evaporating Flare Model}",
      journal = {\solphys},
         year = "1974",
        month = "Feb",
       volume = {34},
       number = {2},
        pages = {323-338},
          doi = {10.1007/BF00153671},
       adsurl = {https://ui.adsabs.harvard.edu/abs/1974SoPh...34..323H}
}

@ARTICLE{kopp1976,
       author = {{Kopp}, R.~A. and {Pneuman}, G.~W.},
        title = "{Magnetic reconnection in the corona and the loop prominence phenomenon.}",
      journal = {\solphys},
         year = "1976",
        month = "Oct",
       volume = {50},
       number = {1},
        pages = {85-98},
          doi = {10.1007/BF00206193},
       adsurl = {https://ui.adsabs.harvard.edu/abs/1976SoPh...50...85K}
}

@ARTICLE{oka2013looptop,
       author = {{Oka}, M. and {Ishikawa}, S. and {Saint-Hilaire}, P. and {Krucker}, S. and {Lin}, R.~P.},
        title = "{Kappa Distribution Model for Hard X-Ray Coronal Sources of Solar Flares}",
      journal = {\apj},
         year = 2013,
        month = feb,
       volume = {764},
       number = {1},
          eid = {6},
        pages = {6},
          doi = {10.1088/0004-637X/764/1/6},
archivePrefix = {arXiv},
       eprint = {1212.2579},
 primaryClass = {astro-ph.SR},
       adsurl = {https://ui.adsabs.harvard.edu/abs/2013ApJ...764....6O}
}

@ARTICLE{warmuth2020,
       author = {{Warmuth}, A. and {Mann}, G.},
        title = "{Thermal-nonthermal energy partition in solar flares derived from X-ray, EUV, and bolometric observations. Discussion of recent studies}",
      journal = {\aap},
         year = 2020,
        month = dec,
       volume = {644},
          eid = {A172},
        pages = {A172},
          doi = {10.1051/0004-6361/202039529},
archivePrefix = {arXiv},
       eprint = {2011.04442},
 primaryClass = {astro-ph.SR},
       adsurl = {https://ui.adsabs.harvard.edu/abs/2020A&A...644A.172W}
}

@ARTICLE{Chen2024,
       author = {{Chen}, Bin and {Kong}, Xiangliang and {Yu}, Sijie and {Shen}, Chengcai and {Li}, Xiaocan and {Guo}, Fan and {Zhang}, Yixian and {Glesener}, Lindsay and {Krucker}, S{\"a}m},
        title = "{Energetic Electrons Accelerated and Trapped in a Magnetic Bottle above a Solar Flare Arcade}",
      journal = {\apj},
         year = 2024,
        month = aug,
       volume = {971},
       number = {1},
          eid = {85},
        pages = {85},
          doi = {10.3847/1538-4357/ad531a},
archivePrefix = {arXiv},
       eprint = {2406.00109},
 primaryClass = {astro-ph.SR},
       adsurl = {https://ui.adsabs.harvard.edu/abs/2024ApJ...971...85C}
}

@ARTICLE{fleishman2022,
       author = {{Fleishman}, Gregory D. and {Nita}, Gelu M. and {Chen}, Bin and {Yu}, Sijie and {Gary}, Dale E.},
        title = "{Solar flare accelerates nearly all electrons in a large coronal volume}",
      journal = {\nat},
         year = 2022,
        month = jun,
       volume = {606},
       number = {7915},
        pages = {674-677},
          doi = {10.1038/s41586-022-04728-8},
       adsurl = {https://ui.adsabs.harvard.edu/abs/2022Natur.606..674F}
}

@article{French2023,
doi = {10.3847/1538-4357/acb7dd},
url = {https://dx.doi.org/10.3847/1538-4357/acb7dd},
year = {2023},
month = {apr},
publisher = {The American Astronomical Society},
volume = {948},
number = {1},
pages = {19},
author = {Omar French and Fan Guo and Qile Zhang and Dmitri A. Uzdensky},
title = {Particle Injection and Nonthermal Particle Acceleration in Relativistic Magnetic Reconnection*},
journal = {The Astrophysical Journal}
}

@ARTICLE{guo2024,
       author = {{Guo}, Fan and {Liu}, Yi-Hsin and {Zenitani}, Seiji and {Hoshino}, Masahiro},
        title = "{Magnetic Reconnection and Associated Particle Acceleration in High-Energy Astrophysics}",
      journal = {\ssr},
         year = 2024,
        month = jun,
       volume = {220},
       number = {4},
          eid = {43},
        pages = {43},
          doi = {10.1007/s11214-024-01073-2},
archivePrefix = {arXiv},
       eprint = {2309.13382},
 primaryClass = {astro-ph.HE},
       adsurl = {https://ui.adsabs.harvard.edu/abs/2024SSRv..220...43G}
}

@ARTICLE{dahlin2022,
       author = {{Dahlin}, Joel T. and {Antiochos}, Spiro K. and {Qiu}, Jiong and {DeVore}, C. Richard},
        title = "{Variability of the Reconnection Guide Field in Solar Flares}",
      journal = {\apj},
         year = 2022,
        month = jun,
       volume = {932},
       number = {2},
          eid = {94},
        pages = {94},
          doi = {10.3847/1538-4357/ac6e3d},
archivePrefix = {arXiv},
       eprint = {2110.04132},
 primaryClass = {astro-ph.SR},
       adsurl = {https://ui.adsabs.harvard.edu/abs/2022ApJ...932...94D}
}

@ARTICLE{seo2024feedback,
       doi = {10.3847/1538-4357/ad8e64},
url = {https://dx.doi.org/10.3847/1538-4357/ad8e64},
year = {2024},
month = {dec},
publisher = {The American Astronomical Society},
volume = {977},
number = {2},
pages = {146},
author = {Seo, Jeongbhin and Guo, Fan and Li, Xiaocan and Li, Hui},
title = {Proton Acceleration in Low-β Magnetic Reconnection with Energetic Particle Feedback},
journal = {The Astrophysical Journal}
}

@ARTICLE{Huang2016,
       author = {{Huang}, Yi-Min and {Bhattacharjee}, A.},
        title = "{Turbulent Magnetohydrodynamic Reconnection Mediated by the Plasmoid Instability}",
      journal = {\apj},
         year = 2016,
        month = feb,
       volume = {818},
       number = {1},
          eid = {20},
        pages = {20},
          doi = {10.3847/0004-637X/818/1/20},
archivePrefix = {arXiv},
       eprint = {1512.01520},
 primaryClass = {physics.plasm-ph},
       adsurl = {https://ui.adsabs.harvard.edu/abs/2016ApJ...818...20H}
}

@ARTICLE{provornikova2016,
       author = {{Provornikova}, E. and {Laming}, J.~M. and {Lukin}, V.~S.},
        title = "{Plasma Compression in Magnetic Reconnection Regions in the Solar Corona}",
      journal = {\apj},
         year = 2016,
        month = jul,
       volume = {825},
       number = {1},
          eid = {55},
        pages = {55},
          doi = {10.3847/0004-637X/825/1/55},
archivePrefix = {arXiv},
       eprint = {1604.07325},
 primaryClass = {physics.space-ph},
       adsurl = {https://ui.adsabs.harvard.edu/abs/2016ApJ...825...55P}
}

@ARTICLE{li2019a,
       author = {{Li}, Xiaocan and {Guo}, Fan and {Li}, Hui},
        title = "{Particle Acceleration in Kinetic Simulations of Nonrelativistic Magnetic Reconnection with Different Ion-Electron Mass Ratios}",
      journal = {\apj},
         year = 2019,
        month = jul,
       volume = {879},
       number = {1},
          eid = {5},
        pages = {5},
          doi = {10.3847/1538-4357/ab223b},
archivePrefix = {arXiv},
       eprint = {1905.08797},
 primaryClass = {physics.plasm-ph},
       adsurl = {https://ui.adsabs.harvard.edu/abs/2019ApJ...879....5L}
}

@ARTICLE{li2018a,
       author = {{Li}, Xiaocan and {Guo}, Fan and {Li}, Hui and {Birn}, Joachim},
        title = "{The Roles of Fluid Compression and Shear in Electron Energization during Magnetic Reconnection}",
      journal = {\apj},
         year = 2018,
        month = mar,
       volume = {855},
       number = {2},
          eid = {80},
        pages = {80},
          doi = {10.3847/1538-4357/aaacd5},
archivePrefix = {arXiv},
       eprint = {1801.02255},
 primaryClass = {physics.plasm-ph},
       adsurl = {https://ui.adsabs.harvard.edu/abs/2018ApJ...855...80L}
}

@ARTICLE{birn2012,
       author = {{Birn}, J. and {Borovsky}, J.~E. and {Hesse}, M.},
        title = "{The role of compressibility in energy release by magnetic reconnection}",
      journal = {Physics of Plasmas},
         year = 2012,
        month = aug,
       volume = {19},
       number = {8},
          eid = {082109},
        pages = {082109},
          doi = {10.1063/1.4742314},
       adsurl = {https://ui.adsabs.harvard.edu/abs/2012PhPl...19h2109B}
}

@ARTICLE{isenberg1979,
       author = {{Isenberg}, P.~A. and {Jokipii}, J.~R.},
        title = "{Gradient and curvature drifts in magnetic fields with arbitrary spatial variation.}",
      journal = {\apj},
         year = 1979,
        month = dec,
       volume = {234},
        pages = {746-752},
          doi = {10.1086/157551},
       adsurl = {https://ui.adsabs.harvard.edu/abs/1979ApJ...234..746I}
}

@ARTICLE{Uzdensky2010,
       author = {{Uzdensky}, D.~A. and {Loureiro}, N.~F. and {Schekochihin}, A.~A.},
        title = "{Fast Magnetic Reconnection in the Plasmoid-Dominated Regime}",
      journal = {\prl},
         year = 2010,
        month = dec,
       volume = {105},
       number = {23},
          eid = {235002},
        pages = {235002},
          doi = {10.1103/PhysRevLett.105.235002},
archivePrefix = {arXiv},
       eprint = {1008.3330},
 primaryClass = {astro-ph.SR},
       adsurl = {https://ui.adsabs.harvard.edu/abs/2010PhRvL.105w5002U}
}

@ARTICLE{Loureiro2016,
       author = {{Loureiro}, N.~F. and {Uzdensky}, D.~A.},
        title = "{Magnetic reconnection: from the Sweet-Parker model to stochastic plasmoid chains}",
      journal = {Plasma Physics and Controlled Fusion},
         year = 2016,
        month = jan,
       volume = {58},
       number = {1},
          eid = {014021},
        pages = {014021},
          doi = {10.1088/0741-3335/58/1/014021},
archivePrefix = {arXiv},
       eprint = {1507.07756},
 primaryClass = {physics.plasm-ph},
       adsurl = {https://ui.adsabs.harvard.edu/abs/2016PPCF...58a4021L}
}

@ARTICLE{kane1983,
       author = {{Kane}, S.~R. and {Kai}, K. and {Kosugi}, T. and {Enome}, S. and {Landecker}, P.~B. and {McKenzie}, D.~L.},
        title = "{Acceleration and confinement of energetic particles in the 1980 June 7 solar flare}",
      journal = {\apj},
         year = 1983,
        month = aug,
       volume = {271},
        pages = {376-387},
          doi = {10.1086/161203},
       adsurl = {https://ui.adsabs.harvard.edu/abs/1983ApJ...271..376K}
}

@ARTICLE{inglis2008,
       author = {{Inglis}, A.~R. and {Nakariakov}, V.~M. and {Melnikov}, V.~F.},
        title = "{Multi-wavelength spatially resolved analysis of quasi-periodic pulsations in a solar flare}",
      journal = {\aap},
         year = 2008,
        month = sep,
       volume = {487},
       number = {3},
        pages = {1147-1153},
          doi = {10.1051/0004-6361:20079323},
       adsurl = {https://ui.adsabs.harvard.edu/abs/2008A&A...487.1147I}
}

@ARTICLE{hayes2020,
       author = {{Hayes}, Laura A. and {Inglis}, Andrew R. and {Christe}, Steven and {Dennis}, Brian and {Gallagher}, Peter T.},
        title = "{Statistical Study of GOES X-Ray Quasi-periodic Pulsations in Solar Flares}",
      journal = {\apj},
         year = 2020,
        month = may,
       volume = {895},
       number = {1},
          eid = {50},
        pages = {50},
          doi = {10.3847/1538-4357/ab8d40},
archivePrefix = {arXiv},
       eprint = {2004.11775},
 primaryClass = {astro-ph.SR},
       adsurl = {https://ui.adsabs.harvard.edu/abs/2020ApJ...895...50H}
}

@article{bykov2024,
  title = {X-ray polarization: A view deep inside cosmic ray driven turbulence and particle acceleration in supernova remnants},
  author = {Bykov, Andrei M. and Osipov, Sergei M. and Uvarov, Yury A. and Ellison, Donald C. and Slane, Patrick},
  journal = {Phys. Rev. D},
  volume = {110},
  issue = {2},
  pages = {023041},
  numpages = {14},
  year = {2024},
  month = {Jul},
  publisher = {American Physical Society},
  doi = {10.1103/PhysRevD.110.023041},
  url = {https://link.aps.org/doi/10.1103/PhysRevD.110.023041}
}

@article{bykov2014,
doi = {10.1088/0004-637X/789/2/137},
url = {https://dx.doi.org/10.1088/0004-637X/789/2/137},
year = {2014},
month = {jun},
publisher = {The American Astronomical Society},
volume = {789},
number = {2},
pages = {137},
author = {Andrei M. Bykov and Donald C. Ellison and Sergei M. Osipov and Andrey E. Vladimirov},
title = {MAGNETIC FIELD AMPLIFICATION IN NONLINEAR DIFFUSIVE SHOCK ACCELERATION INCLUDING RESONANT AND NON-RESONANT COSMIC-RAY DRIVEN INSTABILITIES},
journal = {The Astrophysical Journal}
}

@article{Shen_2018,
doi = {10.3847/1538-4357/aaeed3},
url = {https://dx.doi.org/10.3847/1538-4357/aaeed3},
year = {2018},
month = {dec},
publisher = {The American Astronomical Society},
volume = {869},
number = {2},
pages = {116},
author = {Shen, Chengcai and Kong, Xiangliang and Guo, Fan and Raymond, John C. and Chen, Bin},
title = {The Dynamical Behavior of Reconnection-driven Termination Shocks in Solar Flares: Magnetohydrodynamic Simulations},
journal = {The Astrophysical Journal}
}

@ARTICLE{asai2004,
       author = {{Asai}, Ayumi and {Yokoyama}, Takaaki and {Shimojo}, Masumi and {Shibata}, Kazunari},
        title = "{Downflow Motions Associated with Impulsive Nonthermal Emissions Observed in the 2002 July 23 Solar Flare}",
      journal = {\apjl},
         year = 2004,
        month = apr,
       volume = {605},
       number = {1},
        pages = {L77-L80},
          doi = {10.1086/420768},
       adsurl = {https://ui.adsabs.harvard.edu/abs/2004ApJ...605L..77A}
}

@ARTICLE{Takasao2016,
       author = {{Takasao}, Shinsuke and {Asai}, Ayumi and {Isobe}, Hiroaki and {Shibata}, Kazunari},
        title = "{Observational Evidence of Particle Acceleration Associated with Plasmoid Motions}",
      journal = {\apj},
         year = 2016,
        month = sep,
       volume = {828},
       number = {2},
          eid = {103},
        pages = {103},
          doi = {10.3847/0004-637X/828/2/103},
archivePrefix = {arXiv},
       eprint = {1611.00108},
 primaryClass = {astro-ph.SR},
       adsurl = {https://ui.adsabs.harvard.edu/abs/2016ApJ...828..103T}
}

@ARTICLE{Kong2020,
       author = {{Kong}, Xiangliang and {Guo}, Fan and {Shen}, Chengcai and {Chen}, Bin and {Chen}, Yao and {Giacalone}, Joe},
        title = "{Dynamical Modulation of Solar Flare Electron Acceleration due to Plasmoid-shock Interactions in the Looptop Region}",
      journal = {\apjl},
         year = 2020,
        month = dec,
       volume = {905},
       number = {2},
          eid = {L16},
        pages = {L16},
          doi = {10.3847/2041-8213/abcbf5},
archivePrefix = {arXiv},
       eprint = {2011.09585},
 primaryClass = {astro-ph.SR},
       adsurl = {https://ui.adsabs.harvard.edu/abs/2020ApJ...905L..16K}
}

@ARTICLE{chen2015,
       author = {{Chen}, Bin and {Bastian}, Timothy S. and {Shen}, Chengcai and {Gary}, Dale E. and {Krucker}, S{\"a}m and {Glesener}, Lindsay},
        title = "{Particle acceleration by a solar flare termination shock}",
      journal = {Science},
         year = 2015,
        month = dec,
       volume = {350},
       number = {6265},
        pages = {1238-1242},
          doi = {10.1126/science.aac8467},
archivePrefix = {arXiv},
       eprint = {1512.02237},
 primaryClass = {astro-ph.SR},
       adsurl = {https://ui.adsabs.harvard.edu/abs/2015Sci...350.1238C}
}

@ARTICLE{kontar2023,
       author = {{Kontar}, Eduard P. and {Emslie}, A. Gordon and {Motorina}, Galina G. and {Dennis}, Brian R.},
        title = "{The Efficiency of Electron Acceleration during the Impulsive Phase of a Solar Flare}",
      journal = {\apjl},
         year = 2023,
        month = apr,
       volume = {947},
       number = {1},
          eid = {L13},
        pages = {L13},
          doi = {10.3847/2041-8213/acc9b7},
archivePrefix = {arXiv},
       eprint = {2304.01088},
 primaryClass = {astro-ph.SR},
       adsurl = {https://ui.adsabs.harvard.edu/abs/2023ApJ...947L..13K}
}

@ARTICLE{Takasao2016b,
       author = {{Takasao}, Shinsuke and {Shibata}, Kazunari},
        title = "{Above-the-loop-top Oscillation and Quasi-periodic Coronal Wave Generation in Solar Flares}",
      journal = {\apj},
         year = 2016,
        month = jun,
       volume = {823},
       number = {2},
          eid = {150},
        pages = {150},
          doi = {10.3847/0004-637X/823/2/150},
archivePrefix = {arXiv},
       eprint = {1606.09354},
 primaryClass = {astro-ph.SR},
       adsurl = {https://ui.adsabs.harvard.edu/abs/2016ApJ...823..150T}
}

@ARTICLE{Lin2000,
       author = {{Lin}, Haosheng and {Penn}, Matthew J. and {Tomczyk}, Steven},
        title = "{A New Precise Measurement of the Coronal Magnetic Field Strength}",
      journal = {\apjl},
         year = 2000,
        month = oct,
       volume = {541},
       number = {2},
        pages = {L83-L86},
          doi = {10.1086/312900},
       adsurl = {https://ui.adsabs.harvard.edu/abs/2000ApJ...541L..83L}
}

@ARTICLE{Yang2020,
       author = {{Yang}, Zihao and {Bethge}, Christian and {Tian}, Hui and {Tomczyk}, Steven and {Morton}, Richard and {Del Zanna}, Giulio and {McIntosh}, Scott W. and {Karak}, Bidya Binay and {Gibson}, Sarah and {Samanta}, Tanmoy and {He}, Jiansen and {Chen}, Yajie and {Wang}, Linghua},
        title = "{Global maps of the magnetic field in the solar corona}",
      journal = {Science},
         year = 2020,
        month = aug,
       volume = {369},
       number = {6504},
        pages = {694-697},
          doi = {10.1126/science.abb4462},
archivePrefix = {arXiv},
       eprint = {2008.03136},
 primaryClass = {astro-ph.SR},
       adsurl = {https://ui.adsabs.harvard.edu/abs/2020Sci...369..694Y}
}

@ARTICLE{Brooks2021,
       author = {{Brooks}, David H. and {Warren}, Harry P. and {Landi}, Enrico},
        title = "{Measurements of Coronal Magnetic Field Strengths in Solar Active Region Loops}",
      journal = {\apjl},
         year = 2021,
        month = jul,
       volume = {915},
       number = {1},
          eid = {L24},
        pages = {L24},
          doi = {10.3847/2041-8213/ac0c84},
archivePrefix = {arXiv},
       eprint = {2106.10884},
 primaryClass = {astro-ph.SR},
       adsurl = {https://ui.adsabs.harvard.edu/abs/2021ApJ...915L..24B}
}

@ARTICLE{li2013,
       author = {{Li}, T.~C. and {Drake}, J.~F. and {Swisdak}, M.},
        title = "{Coronal Electron Confinement by Double Layers}",
      journal = {\apj},
         year = 2013,
        month = dec,
       volume = {778},
       number = {2},
          eid = {144},
        pages = {144},
          doi = {10.1088/0004-637X/778/2/144},
archivePrefix = {arXiv},
       eprint = {1402.3915},
 primaryClass = {physics.plasm-ph},
       adsurl = {https://ui.adsabs.harvard.edu/abs/2013ApJ...778..144L}
}

@ARTICLE{chen2020ApJ,
       author = {{Chen}, Bin and {Yu}, Sijie and {Reeves}, Katharine K. and {Gary}, Dale E.},
        title = "{Microwave Spectral Imaging of an Erupting Magnetic Flux Rope: Implications for the Standard Solar Flare Model in Three Dimensions}",
      journal = {\apjl},
         year = 2020,
        month = jun,
       volume = {895},
       number = {2},
          eid = {L50},
        pages = {L50},
          doi = {10.3847/2041-8213/ab901a},
archivePrefix = {arXiv},
       eprint = {2005.01900},
 primaryClass = {astro-ph.SR},
       adsurl = {https://ui.adsabs.harvard.edu/abs/2020ApJ...895L..50C}
}

@ARTICLE{Mondal2024,
       author = {{Mondal}, Surajit and {Battaglia}, Andrea F. and {Chen}, Bin and {Yu}, Sijie},
        title = "{A Joint Microwave and Hard X-Ray Study toward Understanding the Transport of Accelerated Electrons During an Eruptive Solar Flare}",
      journal = {\apj},
         year = 2024,
        month = may,
       volume = {966},
       number = {2},
          eid = {208},
        pages = {208},
          doi = {10.3847/1538-4357/ad3910},
archivePrefix = {arXiv},
       eprint = {2404.14268},
 primaryClass = {astro-ph.SR},
       adsurl = {https://ui.adsabs.harvard.edu/abs/2024ApJ...966..208M}
}

@ARTICLE{Young2009,
       author = {{Young}, P.~R. and {Watanabe}, T. and {Hara}, H. and {Mariska}, J.~T.},
        title = "{High-precision density measurements in the solar corona. I. Analysis methods and results for Fe XII and Fe XIII}",
      journal = {\aap},
         year = 2009,
        month = feb,
       volume = {495},
       number = {2},
        pages = {587-606},
          doi = {10.1051/0004-6361:200810143},
archivePrefix = {arXiv},
       eprint = {0805.0958},
 primaryClass = {astro-ph},
       adsurl = {https://ui.adsabs.harvard.edu/abs/2009A&A...495..587Y}
}

@ARTICLE{DelZanna2023,
       author = {{Del Zanna}, Giulio and {Samra}, Jenna and {Monaghan}, Austin and {Madsen}, Chad and {Bryans}, Paul and {DeLuca}, Edward and {Mason}, Helen and {Berkey}, Ben and {de Wijn}, Alfred and {Rivera}, Yeimy J.},
        title = "{Coronal Densities, Temperatures, and Abundances during the 2019 Total Solar Eclipse: The Role of Multiwavelength Observations in Coronal Plasma Characterization}",
      journal = {\apjs},
         year = 2023,
        month = mar,
       volume = {265},
       number = {1},
          eid = {11},
        pages = {11},
          doi = {10.3847/1538-4365/acad68},
archivePrefix = {arXiv},
       eprint = {2212.11889},
 primaryClass = {astro-ph.SR},
       adsurl = {https://ui.adsabs.harvard.edu/abs/2023ApJS..265...11D}
}

@ARTICLE{Warren2009,
       author = {{Warren}, Harry P. and {Brooks}, David H.},
        title = "{The Temperature and Density Structure of the Solar Corona. I. Observations of the Quiet Sun with the EUV Imaging Spectrometer on Hinode}",
      journal = {\apj},
         year = 2009,
        month = jul,
       volume = {700},
       number = {1},
        pages = {762-773},
          doi = {10.1088/0004-637X/700/1/762},
archivePrefix = {arXiv},
       eprint = {0901.1621},
 primaryClass = {astro-ph.SR},
       adsurl = {https://ui.adsabs.harvard.edu/abs/2009ApJ...700..762W}
}

@ARTICLE{Dahlin2025,
       author = {{Dahlin}, Joel T. and {Antiochos}, Spiro K. and {DeVore}, C. Richard and {Wyper}, Peter F. and {Qiu}, Jiong},
        title = "{Determining the 3D Dynamics of Solar Flare Magnetic Reconnection}",
      journal = {arXiv e-prints},
         year = 2025,
        month = apr,
          eid = {arXiv:2504.00913},
        pages = {arXiv:2504.00913},
          doi = {10.48550/arXiv.2504.00913},
archivePrefix = {arXiv},
       eprint = {2504.00913},
 primaryClass = {astro-ph.SR},
       adsurl = {https://ui.adsabs.harvard.edu/abs/2025arXiv250400913D}
}

@ARTICLE{Ma2023,
       author = {{Ma}, Hanqing and {Drake}, J.~F. and {Swisdak}, M.},
        title = "{Wave Generation and Energetic Electron Scattering in Solar Flares}",
      journal = {\apj},
         year = 2023,
        month = sep,
       volume = {954},
       number = {1},
          eid = {21},
        pages = {21},
          doi = {10.3847/1538-4357/ace59e},
archivePrefix = {arXiv},
       eprint = {2307.04769},
 primaryClass = {astro-ph.SR},
       adsurl = {https://ui.adsabs.harvard.edu/abs/2023ApJ...954...21M}
}

@ARTICLE{Zhang2024ApJ,
       author = {{Zhang}, Qile and {Guo}, Fan and {Daughton}, William and {Li}, Xiaocan and {Li}, Hui},
        title = "{Plasma Dynamics and Nonthermal Particle Acceleration in 3D Nonrelativistic Magnetic Reconnection}",
      journal = {\apj},
         year = 2024,
        month = oct,
       volume = {974},
       number = {1},
          eid = {47},
        pages = {47},
          doi = {10.3847/1538-4357/ad6561},
archivePrefix = {arXiv},
       eprint = {2404.08807},
 primaryClass = {astro-ph.SR},
       adsurl = {https://ui.adsabs.harvard.edu/abs/2024ApJ...974...47Z}
}

@ARTICLE{li2019power-law,
       author = {{Li}, Xiaocan and {Guo}, Fan and {Li}, Hui and {Stanier}, Adam and {Kilian}, Patrick},
        title = "{Formation of Power-law Electron Energy Spectra in Three-dimensional Low-{\ensuremath{\beta}} Magnetic Reconnection}",
      journal = {\apj},
         year = 2019,
        month = oct,
       volume = {884},
       number = {2},
          eid = {118},
        pages = {118},
          doi = {10.3847/1538-4357/ab4268},
archivePrefix = {arXiv},
       eprint = {1909.01911},
 primaryClass = {astro-ph.SR},
       adsurl = {https://ui.adsabs.harvard.edu/abs/2019ApJ...884..118L}
}

@ARTICLE{Yokoyama2001,
       author = {{Yokoyama}, Takaaki and {Shibata}, Kazunari},
        title = "{Magnetohydrodynamic Simulation of a Solar Flare with Chromospheric Evaporation Effect Based on the Magnetic Reconnection Model}",
      journal = {\apj},
         year = 2001,
        month = mar,
       volume = {549},
       number = {2},
        pages = {1160-1174},
          doi = {10.1086/319440},
       adsurl = {https://ui.adsabs.harvard.edu/abs/2001ApJ...549.1160Y}
}

@ARTICLE{Drury1983,
       author = {{Drury}, L. Oc.},
        title = "{REVIEW ARTICLE: An introduction to the theory of diffusive shock acceleration of energetic particles in tenuous plasmas}",
      journal = {Reports on Progress in Physics},
         year = 1983,
        month = aug,
       volume = {46},
       number = {8},
        pages = {973-1027},
          doi = {10.1088/0034-4885/46/8/002},
       adsurl = {https://ui.adsabs.harvard.edu/abs/1983RPPh...46..973D}
}

@ARTICLE{Seaton2009,
       author = {{Seaton}, Daniel B. and {Forbes}, Terry G.},
        title = "{An Analytical Model for Reconnection Outflow Jets Including Thermal Conduction}",
      journal = {\apj},
         year = 2009,
        month = aug,
       volume = {701},
       number = {1},
        pages = {348-359},
          doi = {10.1088/0004-637X/701/1/348},
       adsurl = {https://ui.adsabs.harvard.edu/abs/2009ApJ...701..348S}
}

@ARTICLE{Yokoyama1998,
       author = {{Yokoyama}, Takaaki and {Shibata}, Kazunari},
        title = "{A Two-dimensional Magnetohydrodynamic Simulation of Chromospheric Evaporation in a Solar Flare Based on a Magnetic Reconnection Model}",
      journal = {\apjl},
         year = 1998,
        month = feb,
       volume = {494},
       number = {1},
        pages = {L113-L116},
          doi = {10.1086/311174},
       adsurl = {https://ui.adsabs.harvard.edu/abs/1998ApJ...494L.113Y}
}
\bibliographystyle{aasjournal}

\end{document}